\documentclass{article}
\usepackage{latexsym}
\newtheorem{theorem}{Theorem}
\newtheorem{lemma}{Lemma}
\newtheorem{definition}{Definition}
 \def \dsl{\raise .15ex\hbox{/}\kern-.57em\partial}
 \def \Dsl{\,\raise .15ex\hbox{/}\mkern-13.5mu D}

\begin{document}
\title{Short Distance Asymptotics of Ising Correlations}
\author{John Palmer \\\ Department of Mathematics\\
University of Arizona\\
Tucson, AZ 85721}
\maketitle
\begin{abstract}
We prove that the short distance asymptotics for the
even Ising model scaling functions from below
$T_c$ is given by the Luther-Peschel formula. Generalizations
to the odd scaling functions and Holonomic Fields are
given.
\end{abstract}
\section{Introduction}
In this paper we will use the Sato, Miwa, Jimbo 
characterization of the scaling functions for the two 
dimensional Ising model to show that the short distance
asymptotics of the even scaling functions below the critical
point are given by the Luther-Peschel formula (see 
Theorem(\ref{main}) below).  We will then present
results for the odd correlations below $T_c$ and also for
holonomic quantum fields which are a consequence of the
same technique used to prove Theorem(\ref{main}).

This paper is a sequel to \cite{P00} and the reader is
referred to that paper for a more detailed explanation 
of the Ising model scaling limits than we will give here.
Continuum limits for the two dimensional Ising 
correlations on a lattice were first considered in 
\cite{WMTB76}, where, in addition, a connection with Painlev\'e 
transcendents was discovered.  In a series of papers
Sato, Miwa, and Jimbo showed that the continuum
correlations (the scaling functions) were associated with
monodromy preserving deformations of the Euclidean
Dirac equation and that this connection sufficed to 
account for the appearance of the Painlev\'e 
transcendents,
\cite{SMJ78}-\cite{SMJ80}.  Here we
exploit the fact that the
SMJ formula for the log derivative of the scaling 
function (a $\tau$ function in their terminology) can be
expressed in terms of the Fourier coefficients of a 
solution to the {\em linear\/} Dirac equation.  We analyse the
linear problem in order to control the 
short distance asymptotics.  This analysis was suggested
by the success of Riemann-Hilbert techniques in obtaining
asymptotics for non linear integrable systems 
\cite{DIZ97}, where a similar connection with a linear 
problem is a central feature. 

We would like to point out that the two point function
both for the Ising model and for Holonomic Fields in
general has been analysed in more detail than the
result we obtain here, \cite{T91}, \cite{BT92}, and 
\cite{TW98}.  In particular, the constant
term in the short distance asymptotics is obtained--our result
for the log derivative has nothing to say about this.

We will begin by recalling some of the results 
of \cite{P00} where a sketch of the the proof was presented. 
The SMJ characterization involves certain solutions to
the Dirac equation in two dimensions so we will start
with a description of the situation of interest to us.
The Euclidean Dirac operator in ${\bf R}^2$ (with a mass 
perturbation) is given by
\[mI-\dsl=\left[\begin{array}{c}
\begin{array}{cc}
m&-2\partial\\
-2\bar{\partial}&m\end{array}
\end{array}
\right],\]
where,
\begin{eqnarray*}
\partial :=\frac 12\left(\frac {\partial}{\partial x_1}-i\frac {\partial}{
\partial x_2}\right),\\
\bar{\partial }:=\frac 12\left(\frac {\partial}{\partial x_1}+i\frac {
\partial}{\partial x_2}\right).\end{eqnarray*}
Although we will {\em not\/} be working exclusively with 
holomorphic functions, the presence of $\partial$ and $\bar{\partial}$ in
the Dirac operator makes it very convenient to 
introduce the complex variable $z=x_1+ix_2$ with
$\bar {z}=x_1-ix_2$; we thus identify ${\bf R}^2$ with ${\bf C}$ in the
usual fashion.  For brevity we will write $f(z)$ for a 
function of two real variables even though it is 
customary to use a notation like $f(z,\bar {z})$ to avoid the 
temptation to regard $f(z)$ as a holomorphic function
of $z$.

Let ${\bf a}=\{a_1,a_2,\ldots ,a_N\}$ denote a collection of $N$ distinct
points in ${\bf C}$.  The solutions of the Dirac equation that
we are interested in are smooth sections of a rank
2 vector bundle over the punctured plane ${\bf C}\backslash {\bf a}$. 

For
the purpose of allowing some later remarks we will
begin by defining a slightly more general family of {\em line\/} bundles, 
${\cal E}_{\lambda}$, 
than is relevant for the Ising model.  For $j=1,2,\ldots ,N$ suppose
{\em real\/} numbers $\lambda_j$ are given with $|\lambda_j|\le\frac 
12$.  Define
\[\Lambda_j=e^{2\pi i\lambda_j},\]
and write,
\[\lambda =(\lambda_1,\lambda_2,\ldots ,\lambda_N).\]

Roughly speaking the smooth sections of the bundle 
${\cal E}_{\lambda}\rightarrow {\bf C}\backslash {\bf a}$ will be multivalued functions on $
{\bf C}\backslash {\bf a}$ with
values in ${\bf C}$ which
have multiplier $\Lambda_j$ when continued about a loop
that circles $a_j$ counterclockwise.  This can be made 
precise in an 
elegant fashion by
working on the simply connected covering space of
${\bf C}\backslash {\bf a}$ and then restricting attention to smooth sections that
transform appropriately under the action of 
$\pi_1({\bf C}\backslash {\bf a})$ by deck transformations.  However, some later
developments will be clearer for us if we can use 
functions with specific branching behavior as multipliers
taking sections of ${\cal E}_{\lambda}$ to sections of the trivial bundle
over ${\bf C}\backslash {\bf a}$.  It will be easiest to be precise about this 
multiplier action if we define the
bundles ${\cal E}_{\lambda}$ by giving transition functions, in spite of the 
fact that this is a little clumsy.  

To begin, 
note that there are only a finite number of lines each
of which
consists of all multiples of $a_i-a_j$ for $i$ and $j$ distinct. 
Thus it is possible to choose a vector ${\bf r}\ne 0$ which is not 
contained in any of these
lines.  Then the rays, ${\bf r}_j$, defined by
\[{\bf r}_j=\{z:z=a_j+t{\bf r},t>0\},\]
do not intersect. Choose an argument $\theta_{{\bf r}}$ for ${\bf r}$ so that
${\bf r}=|{\bf r}|e^{i\theta_{{\bf r}}}$ with $|\theta_{{\bf r}}|
\le\pi$ and let $\theta (z)$ denote the polar angle with
\begin{eqnarray*}
\theta_{{\bf r}}-\pi <\theta (z)<\theta_{{\bf r}}+\pi\\
z=|z|e^{i\theta (z)},\end{eqnarray*}
This angle is branched along the ray $-{\bf r}$.
 For $\epsilon >0$ define a tubular 
neighborhood, ${\bf t}_j(\epsilon )$, of ${\bf r}_j$ by,
\[{\bf t}_j(\epsilon )=\{z:\mbox{\rm dist}(z,{\bf r}_j)<\epsilon 
\}\cap \{z:|\theta (z_j)-\theta_{{\bf r}}|<\frac {\pi}4\}.\]
Now choose 
$\epsilon >0$ small enough so that the tubular neighborhoods ${\bf t}_
j(\epsilon )$ 
are mutually disjoint {\em and\/} so that the disks,
\[D_j(2\epsilon ):=\{z:|z-a_j|<2\epsilon \},\]
are also mutually disjoint (this will be useful later on).

We now introduce a covering of ${\bf C}\backslash {\bf a}$ over each element 
of which the bundle ${\cal E}_{\lambda}$ is trivial.  Let
\[{\cal U}_0:=\{z\in {\bf C}\backslash {\bf a}:z\notin {\bf r}_j\mbox{\rm \ for }
j=1,2,\ldots ,N\},\]
 Let
\[{\cal U}_j:={\bf t}_j(\epsilon )\mbox{\rm \ for }j=1,2,\ldots ,
N.\]
Now we glue together the trivial bundles,
\[{\cal U}_k\times {\bf C}\rightarrow {\cal U}_k\mbox{\rm \ for }
k=0,1,\ldots ,N,\]
by giving the transition functions $s_j$ that define the
bundle ${\cal E}_{\lambda}$.  For $j=1,\ldots ,N$ define,
\[s_j(z)=\left\{\begin{array}{c}
\Lambda_j\mbox{\rm \ for }\theta (z_j)<0\\
1\mbox{\rm \ for }\theta (z_j)>0.\end{array}
\right.\]
Then the bundle ${\cal E}_{\lambda}$ is defined by the following transition
maps between vectors $(z,v)_0\in {\cal U}_0\times {\bf C}$ in the trivial 
bundle
over ${\cal U}_0$ and vectors $(z,v)_j\in {\cal U}_j\times {\bf C}$ in the trivial bundle
over ${\cal U}_j$ (for $k=1,2,\ldots ,N),$
\[(z,v)_0=(z,s_j(z)v)_j\mbox{\rm \ for }z\in {\cal U}_0\cap {\cal U}_
j.\]
The function $s_j(z)$ is smooth since it is constant on each
of the two components of ${\cal U}_0\cap {\cal U}_j$.  The bundle that is 
relevant for the Ising model is the one with the choice
$\Lambda_j=-1$ for all $j=1,2,\ldots ,N$.  For simplicity we will 
denote this bundle by ${\cal E}$ with no subscript. 

The rank 2 {\em vector\/} bundles that are more directly of 
interest to us are ${\cal E}_{\lambda}\otimes {\bf C}^2$ and ${\cal E}
\otimes {\bf C}^2$, the direct
sum of two copies of ${\cal E}_{\lambda}$ and ${\cal E}$ respectively.  For
simplicity we will use the same notation, ${\cal E}_{\lambda}$ and $
{\cal E}$, 
to denote these
vector bundles and when necessary make distinctions
by referring to the {\em line bundles} ${\cal E}_{\lambda}$ and $
{\cal E}$.

The differential operator $mI-\dsl$ acts on $C^{\infty}({\cal E}_{
\lambda})$, the space of
smooth sections of the vector bundle ${\cal E}_{\lambda}$, since it 
commutes with multiplication by constants.  We will 
now define a family of {\em local\/} smooth sections of 
$C^{\infty}({\cal E}_{\lambda})$ which are simultaneously solutions of the 
Dirac equation, $(mI-\dsl)w=0$ and eigenfunctions for the
infinitesimal rotation about $a_j$, 
$R_j=z_j\partial_j-\bar {z}_j\bar{\partial}_j+\frac 12\left(\begin{array}{cc}
1&0\\
0&-1\end{array}
\right)$, which
commutes with $mI-\dsl$. 
We write $z_j=z-a_j$ and 
$\partial_j=\partial_{z_j}$.  
 Note that this infinitesimal 
rotation has eigenvalues which are translated by $\pm\frac 12$ 
compared to the infinitesimal monodromy.  Following
SMJ we will parametrize our local wave functions
by the $R_j$ eigenvalue rather than the infinitesimal
monodromy.

Let $\Theta (z)$ denote the angular coordinate at $0$
defined so that for $z\notin \{t{\bf r}:t>0\}$ we have,
\[\begin{array}{rcl}
z=|z|e^{i\Theta (z)},\mbox{\rm \ with }\theta_{{\bf r}}<\Theta (z
)<\theta_{{\bf r}}+2\pi\end{array}
.\]

For $\ell$ a real number we define a function $w_{\ell}(z)$ for $
z\in {\bf C}\backslash {\bf r}$ by,
\[w_{\ell}^0(z)=\left(\begin{array}{c}
e^{i(\ell -\frac 12)\Theta (z)}I_{\ell -\frac 12}(m|z|)\\
e^{i(\ell +\frac 12)\Theta (z_{})}I_{\ell +\frac 12}(m|z|)\end{array}
\right),\]
where $I_k$ is the modified Bessel function of order $k$.  
For $z\in {\bf C}\backslash (-{\bf r})$ we define,
\[w^{\pi}_{\ell}(z)=\left(\begin{array}{c}
e^{i(\ell -\frac 12)\theta (z)}I_{\ell -\frac 12}(m|z|)\\
e^{i(\ell +\frac 12)\theta (z_{})}I_{\ell +\frac 12}(m|z|)\end{array}
\right),\]
The only difference being, of course, the choice of angle.  
Where defined these are solutions to the Dirac equation
$(mI-\dsl)w=0$ and are eigenfunctions of the infinitesimal
rotation $Rw_{\ell}=\ell w_{\ell}$ about 0\cite{SMJ79b}.  Now let $
\ell$ denote a real
number and
define (for $|z_j|<2\epsilon$ say),
\begin{eqnarray*}
w_{\ell}(z_j)=w_{\ell}^0(z_j)\mbox{\rm \ in the }{\cal U}_0\mbox{\rm \ trivialization}\\
w_{\ell}(z_j)=w_{\ell}^{\pi}(z_j)\mbox{\rm \ in the }{\cal U}_j\mbox{\rm \ trivialization}\end{eqnarray*}
Then it is easy to check that $w_{\ell}(z_j)$ is a {\em local\/} section
of $C^{\infty}({\cal E}_{\lambda})$ provided $\ell\equiv\frac 12+
\lambda_j\mbox{\rm mod }{\bf Z}$.  Now define
a conjugation on ${\bf C}^2$ by,
\[\left(\begin{array}{c}
a\\
b\end{array}
\right)^{*}=\left(\begin{array}{c}
\bar {b}\\
\bar {a}\end{array}
\right).\]
This conjugation commutes with the Dirac operator $\dsl$
and we define,
\[w^{*}_{\ell}(z)=\left(\begin{array}{c}
\bar {w}_{\ell ,2}(z)\\
\bar {w}_{\ell ,1}(z)\end{array}
\right).\]
One can check that $w_{\ell}^{*}(z_j)$ is a local smooth section of 
$C^{\infty}({\cal E}_{\lambda})$ if and only if $\ell\equiv\frac 
12-\lambda_j\mbox{\rm mod }{\bf Z}$.  It is a result
of SMJ that every solution to $(mI-\dsl)w=0$ in 
$C^{\infty}({\cal E}_{\lambda})$ has local expansions,
\begin{equation}w(z)=\sum_{k\in {\bf Z}+\frac 12}a_k^j(w)w_{k+\lambda_
j}(z_j)+b_k^j(w)w^{*}_{k-\lambda_j}(z_j),\label{lexp}\end{equation}
valid for $0<|z_j|<2\epsilon$ \cite{SMJ79b}, \cite{P93}.  As the reader may check 
the coefficients $a_k^j(w)$ and 
$b_k^j(w)$ are simply related to Fourier coefficients in the
expansion of the restriction of $w$ to say 
the circle $|z_j|=\epsilon$. We will refer to these coefficients as
local expansion coefficients.

For the Ising case $\lambda_j=\pm$$\frac 12$ and
it is better not to use this form of the expansion (which
would require a somewhat arbitrary choice of $\pm\frac 12$ at each $
a_j$); 
instead 
we will just write,
\begin{equation}w(z)=\sum_{n\in {\bf Z}}c_n^j(w)w_n(z_j)+c_n^{j*}
(w)w_n^{*}(z_j).\label{lexpi}\end{equation}
Note that we have changed the names of the local
expansion coefficients in (\ref{lexpi}) to $c_n^j(w)$ and $c_n^{j
*}(w)$ 
so that it  
coincides with the terminology in \cite{SMJ79b}. Our way of writing
(\ref{lexp}) is different than the corresponding local 
expansions in \cite{SMJ79b} and so we have given different names to
the local expansion coefficients. 

Before we move on we will make one further 
observation about local expansions in a neighborhood
of $\infty$.  Suppose that 
$R>0$ is big enough so that all the points $a_j$ for 
$j=1,2,\ldots ,N$ are inside the circle of radius $R$.  Then
$\{z:|z|>R\}\backslash\cup_j{\bf r}_j$ splits into $N$ distinct components and
the ${\cal U}_0$ trivialization is not very convenient for the
description of sections of ${\cal E}$ over this set.  In particular
suppose that $N$ is even.  Then we can alternately flip 
the
signs of sections supported in adjacent components 
of the ${\cal U}_0$ trivialization to
produce a trivialization ${\cal U}_{\infty}$ for ${\cal E}$ over $
\{z:|z|>R\}$.  
Actually the ${\cal U}_0$ trivialization is not defined over
the rays ${\bf r}_j$ but because of the sign flips on adjacent
components it is easy to see that ${\cal U}_{\infty}$ extends to
a trivialization of ${\cal E}$ over the exterior of the disk
of radius $R$.  It is also clear that ${\cal U}_{\infty}$ is only determined
itself up to an overall sign which we fix  by
declaring the ${\cal U}_{\infty}$ trivialization of the ${\cal U}_
0$ section
$\prod_j(z-a_j)^{\epsilon_j}$ for $|\epsilon |=0$ to be,
\[\prod_j\left(1-\frac {a_j}z\right)^{\epsilon_j}\mbox{\rm \ for }
|z|>R,\]
where the fractional powers in this last product are
the holomorphic functions of $z$ normalized to be 1 at $z=\infty$.

 It can be shown (\cite{SMJ79b}, \cite{P93}) that sections
$w\in L^2({\cal E})$ which are solutions to the Dirac equation
in the exterior of the disk of radius $R$ have
convergent expansions (in the ${\cal U}_{\infty}$ trivialization),
\begin{equation}w(z)=\sum_{n\in {\bf Z}}c_n^{\infty}(w)\hat {w}_n
(z),\label{expinf}\end{equation}
where, 
\[\hat {w}_n(z):=\left[\begin{array}{c}
-e^{-in\theta}K_n(m|z|)\\
e^{-i(n-1)\theta}K_{n-1}(m|z|)\end{array}
\right].\]
The functions $K_n$ are the modified Bessel functions
that tend to zero at $\infty$.  The reader should note that
there is more than one definition of these functions
(differing by a factor $e^{in\pi}$).  We are using the version
defined in \cite{O74}.
Also note that because $n$ is an integer
the choice of angle $\theta$ is irrelevant.

Now write $x\cdot y=x_1y_1+x_2y_2$ for the standard bilinear 
form on ${\bf C}^2$, so that $\bar {x}\cdot y$ is the standard Hermitian 
form.  For $w,v\in C_0^{\infty}({\cal E}_{\lambda})$ define an inner product,
\[(w,v)=\frac i2\int_{{\bf C}}\bar {w}\cdot v\,dzd\bar {z},\]
which is well defined since $\bar {w}(z)\cdot v(z)$ decends to a 
compactly supported function on ${\bf C}\backslash {\bf a}$. 
We will write $L^2({\cal E}_{\lambda})$ for the Hilbert space 
completion of $C_0^{\infty}({\cal E}_{\lambda})$ with respect to the norm
induced by this inner product. 

For the rest of this introduction we will specialize our
considerations to the situation relevant to the Ising 
model.  For $n$ an integer we write,
\begin{eqnarray*}
w_n^R=\frac 12(w_n+w_n^{*}),\\
w_n^I=\frac 1{2i}(w_n-w_n^{*}),\end{eqnarray*}
for the real and imaginary parts of $w_n$ with respect to
the conjugation $*$.  Since $\Lambda_j=-1$ is real for all $j$ it follows that
$w_n^R(z_j)$ and $w_n^I(z_j)$ are local sections of $C^{\infty}({\cal E}
)$.
In \cite{P00} it is shown that for $j=1,2,\cdots ,N$ there exists a real 
solution ${\cal W}_j$ $({\cal W}_j^{*}={\cal W}_j)$ to the Dirac equation,
\[(mI-\dsl){\cal W}_j=0,\]
which is in $L^2({\cal E})$ and which has leading order local 
expansions given by,
\begin{equation}{\cal W}_j(z)=\delta_{ij}w_0^I(z_i)+T_{ij}w_0^R(z_
i)+\cdots\mbox{\rm \ for }i=1,2,\ldots ,N\label{rexp}\end{equation}
Note that the coefficients $w_n(z)$ are less and less locally
singular at $z=0$ as $n$ increases.  The $+\cdots$ in (\ref{rexp})
refer to terms with $w_n$ and $w_n^{*}$ for $n>0.$  Also note that
in (\ref{rexp}) it is not necessary to specify what the 
coefficients $T_{ij}$ are--they are already uniquely determined by
the other conditions on ${\cal W}_j$ \cite{P00}.  

We are now ready to present 
the SMJ characterization of the Ising model scaling 
function from below $T_c$, $\tau_{-}(ma)=\tau_{-}(ma_1,ma_2,\ldots 
,ma_N)$.  It is,
\begin{equation}d_a\log\tau_{-}(ma)=\frac m{2i}\sum_jc^j_1({\cal W}_
j)da_j-\overline {c_1^j({\cal W}_j)}d\bar {a}_j.\label{smjchr}\end{equation}
The reader might want to consult \cite{PT83} or 
\cite{P00} for an 
explanation of what exactly $\tau_{-}$ is and how it is related
to two dimensional Ising correlations.
Most of the rest of this paper will be devoted to 
understanding the solution ${\cal W}_j$ well enough in the limit
$m\rightarrow 0$ so that we can compute the limiting values of the
coefficients $mc^j_1({\cal W}_j)$ which appear in (\ref{smjchr}).
Our principal result is,
\begin{theorem}[Luther-Peschel Asymptotics]\label{main}
Suppose that $N$ is even.  Then
\begin{equation}\lim_{m\rightarrow 0}d_a\log\tau_{-}(ma)=\frac 12
d_a\log\sum_{|\epsilon |=0}\prod_{i<j}|a_i-a_j|^{2\epsilon_i\epsilon_
j}\label{LPA}\end{equation}
where the sum is over all choices of $\epsilon_k=\pm\frac 12$ with,
\[|\epsilon |:=\epsilon_1+\epsilon_2+\cdots +\epsilon_N=0.\]
\end{theorem}
After the proof of this result we will indicate the 
changes that are needed to adapt the proof to the case
where $N$ is odd.  We find for $N$ odd,
\[\lim_{m\rightarrow 0}d_a\log\tau_{-}(ma)=\frac 12d_a\log\sum_{|
\epsilon |=\pm\frac 12}\prod_{i<j}|a_i-a_j|^{2\epsilon_i\epsilon_
j}.\]
We will also indicate how to derive the short distance 
behavior of the correlations for Holonomic Fields.

Very briefly the rest of the paper is organized as 
follows.  In section 2 we characterize ${\cal W}_j$ as the
solution to a boundary value problem on a finite
domain.  In the third section we introduce the Green
function for the $m\rightarrow 0$ limit of this boundary value
problem.  In the fourth section we introduce the 
associated boundary value projection.  In the fifth
section we discuss the inversion of a suitable restriction
of this projection.  In the sixth section we discuss
how to put these results together to give 
the perturbation scheme which we use to approximate
${\cal W}_j$ in the limit $m\rightarrow 0$.  In
the seventh section we examine the application of the
same technique to other problems. 
\section{An Equivalent Boundary Value Problem}
The tool we will use in dealing with the $m\rightarrow 0$ limit
of ${\cal W}_j$ is a characterization of ${\cal W}_j$ as the solution to an 
inhomogeneous boundary value problem.  We will now 
describe this characterization.  It is a result of 
SMJ\cite{SMJ79b}
that the space of solutions $w\in C^{\infty}({\cal E})$ to the Dirac
equation,
\[(mI-\dsl)w=0,\]
which are also in $L^2({\cal E})$ is $N$ dimensional. We write
{\bf N} for this space of solutions.  For $w\in {\bf N}$ define,
\[c_0(w)=(c_0^1(w),c_0^2(w),\ldots ,c_0^N(w))\in {\bf C}^N,\]
with a similar definition for $c_0^{*}(w)$.  Now let ${\cal N}$ denote
the image of ${\bf N}$ in ${\bf C}^N\oplus {\bf C}^N$  under the map,
\[{\bf N}\ni w\rightarrow (c_0(w),c_0^{*}(w)).\]
Suppose now that ${\cal I}$ is any subspace of ${\bf C}^N\oplus {\bf C}^
N$
which is {\em transverse\/} to ${\cal N}$. If $f\in C_0^{\infty}(
{\cal E})$ then in \cite{P00} it 
was proved that there exists a unique solution
$w\in L^2({\cal E})$ to
\[(mI-\dsl)w=f\]
which satisfies the boundary condition $(c_0(w),c_0^{*}(w))\in {\cal I}$.
It was also shown there that the subspace ${\cal I}$ given
by the set of vectors $(v,v)$ for $v\in {\bf C}^N$ (the diagonal 
subspace) is transverse to ${\cal N}$.  Henceforth we will work 
with the subspace
${\cal I}$  which 
corresponds to the boundary condition,
\begin{equation}c_0(w)=c_0^{*}(w).\label{bndry}\end{equation}
Now we will make a subtraction from ${\cal W}_j$ which will
put the result in the subspace of sections of ${\cal E}$ 
satisfying (\ref{bndry}).  Let $\varphi (z)$ denote a non negative 
function in
$C^{\infty}_0({\bf R}^2)$ which is identically 1 for $|z|<1$ and identically
0 outside the ball of radius 2.  Define,
\[\varphi_{j,\epsilon}(z)=\varphi\left(\frac {z-a_j}{\epsilon}\right
).\]
Then since $\epsilon$ has been chosen small enough we know that
$\varphi_{j,\epsilon}$ is one in a neighborhood of $a_j$ and vanishes near $
a_i$ 
for all $i\ne j$.  Now define,
\begin{equation}\delta {\cal W}_j(z)=m^{\frac 12}({\cal W}_j(z)-\varphi_{
j,\epsilon}(z)w_0^I(z-a_j)).\label{modw}\end{equation}
Then consulting (\ref{rexp}) we see that if we look at 
the local expansion for $\delta {\cal W}_j$ in an $\epsilon$ neighborhood of $
a_j$ 
then the local expansion coefficients satisfy the 
condition (\ref{bndry}).  The scale factor $m^{\frac 12}$ has been 
introduced so that the following limit exists,
\[\lim_{m\rightarrow 0}m^{\frac 12}w_0^I(z_j)=\frac 1{\sqrt {2\pi}
i}\left[\begin{array}{c}
z_j^{-\frac 12}\\
-\bar {z}_j^{-\frac 12}\end{array}
\right].\]
Here we used $\Gamma (\frac 12)=\sqrt {\pi}$ , the fractional powers of $
z_j$ and
$\bar {z}_j$
that occur are branched along $z={\bf r}_j$ 
and we employ the convention that the section $w_0^I$ can
be identified with its ${\cal U}_0$ trivialization (which appears
on the right hand side).  Using the fact that both
${\cal W}_j$ and $w_0^I(z_j)$ satisfy the massive Dirac equation
we find that,
\begin{equation}(m-\dsl)\delta {\cal W}_j=\left[\begin{array}{cc}
0&-2\partial\varphi_j\\
-2\bar{\partial}\varphi_j&0\end{array}
\right]m^{\frac 12}w_0^I(z_j):=f_j.\label{inhom}\end{equation}
We are now prepared to give an alternative 
characterization of $\delta {\cal W}_j$.  Choose $R>0$ big enough so
that $D_i(2\epsilon )$ is contained inside $|z|=R$ for $i=1,2,\ldots 
,N.$
Let $D_{\infty}=\{z:|z|\ge R\}$ and define the bounded domain, $\Omega$,
by,
\[\Omega ={\bf C}\backslash\left\{\cup_{i=1}^nD_i(\epsilon )\cup 
D_{\infty}\right\}.\]
We write $H^k({\cal E}_{\Omega})$ for the Sobolev space of sections 
of ${\cal E}$ over $\Omega$ which are in $L^2(\Omega )$ together with all their
weak derivatives up to and including those of order $k.$
\begin{lemma}\label{pchar} 
The smooth section $\delta {\cal W}_j$ of ${\cal E}_{\Omega}$ (the restriction of $
{\cal E}$ to
$\Omega$) is uniquely characterized by the following three 
properties,
\begin{list}{}{\setlength{\leftmargin}{.1in}}
\item{\bf 1.} $\delta W_j\in H^1({\cal E}_{\Omega})$ satisfies the 
inhomogeneous Dirac equation (\ref{inhom}) in $\Omega$.
\item{\bf 2.} For $i=1,2,\ldots ,N$ the local Fourier expansions 
(\ref{lexpi}) for
$\delta {\cal W}_j$ restricted to $C_{\epsilon}(a_i)$ have
coefficients $c_k^i(\delta {\cal W}_j)$ and $c_k^i{}^{*}(\delta {\cal W}_
j)$ that vanish for 
$k<0$ and are equal for $k=0$.
\item{\bf 3.} The section $\delta {\cal W}_j$ has a Fourier
expansion $\delta {\cal W}_j=\sum_{n\in {\bf Z}}c_n^{\infty}(\delta 
{\cal W}_j)\hat {w}_n(z)$, on the circle of radius $R$.
\end{list}

\end{lemma}

\emph{Remark:} Henceforth we we interpret the 
local expansions (\ref{lexpi})
as the Fourier expansions of the
restrictions of the ${\cal U}_0$ trivialization of $\delta {\cal W}_
j$ to $|z_i|=\epsilon$
in powers $e^{i(n+\frac 12)\Theta_i}$ for $n\in {\bf Z}$.  In a similar fashion
we interpret (\ref{expinf})
as the Fourier expansions 
of the ${\cal U}_{\infty}$ trivialization of $\delta {\cal W}_j$ restricted to the circle
of radius $R$. 

\emph{Proof of Lemma (\ref{pchar}).} Because the 
solution, $\delta {\cal W}_j$, of condition (1) of 
the Lemma is assumed to be in $H^1({\cal E}_{\Omega})$ it follows
from local elliptic regularity that the solution is actually
in $C^{\infty}({\cal E}_{\Omega})$. The support properties of the inhomogeneous 
term $f_j$ makes it possible to enlarge each circle $C_{\epsilon}
(a_i)$ 
to an annular region in which $\delta {\cal W}_j$ satisfies the 
homogeneous Dirac equation.  In this region it will have
a convergent local expansion of type (\ref{lexpi}).  Since
the Fourier coefficients 
$c_k^i(\delta {\cal W}_j)$ and $c_k^{i*}(\delta {\cal W}_j)$ 
(for the restriction of $\delta {\cal W}_j$ to 
$C_{\epsilon}(a_i)$) 
vanish for
$k<0$ and are equal for $k=0$ it follows (by the 
uniqueness of Fourier expansions) that the same is
true for the local expansion coefficients in the annulus.
Since the Bessel functions $I_{\ell}(r)$ are monotone increasing
functions of $r$
for $\ell\ge 0$ this restriction on the local expansions implies
that they converge in a domain $0<|z_i|<\epsilon'$ where $\epsilon'$ is
slightly bigger than $\epsilon$ (only a finite number of Fourier
coefficients will get larger for smaller values of $|z_j|$).  
This shows that a solution, $\delta {\cal W}_j,$ to
(1) and (2) of the Lemma extends to a solution of the
Dirac equation which is in $L^2$ near $a_i$ and has 
appropriate restrictions on its local expansion 
coefficients.  The same sort of argument shows that
the restriction (3) allows one to extend $\delta {\cal W}_j$ to an
$L^2$ solution to the Dirac equation in a neighborhood of
$\infty$. QED

Without much difficulty the reader should
be able to verify the following formula for the local
expansion coefficients $mc_1^j({\cal W}_j)$ that appear in the SMJ
formula for the log derivative of the $\tau$ function,
\begin{equation}mc_1^j({\cal W}_j)=\frac {\sqrt {m}}{2\pi I_{\frac 
12}(m\epsilon )}\int_{\theta_{{\bf r}}}^{\theta_{{\bf r}}+2\pi}(\delta 
{\cal W}_j)_1(\epsilon e^{i\Theta_j})e^{-i\frac {\Theta_j}2}d\Theta_
j.\label{fcoef}\end{equation}
In this formula $(\delta {\cal W}_j)_1$ is the first component of $
\delta {\cal W}_j$ in 
the ${\cal U}_0$ trivialization.  The formula follows easily from 
the standard formula for Fourier coefficients and the 
fact that the subtraction
of $\varphi_jw_0(z_j)$ does not alter the local expansion 
coefficients at level 1, so that $c_1^j(\delta {\cal W}_j)=\sqrt 
mc_1^j({\cal W}_j)$.

Our strategy in controlling the $m\rightarrow 0$ limit of the 
coefficients $mc_1^j({\cal W}_j)$ will be to use the characterization
of Lemma(\ref{pchar}) in conjunction with the formula
(\ref{fcoef}).  Since, 
\[\lim_{m\rightarrow 0}\frac {\sqrt m}{I_{\frac 12}(m\epsilon )}=\sqrt {\frac 
2{\epsilon}}\Gamma\left(\frac 32\right)=\sqrt {\frac {\pi}{2\epsilon}}
,\]
it will suffice for our purposes to control the 
$m\rightarrow 0$
convergence of $\delta {\cal W}_j$ in $L^p(C_{\epsilon}(a_i))$ for any $
p\ge 1$ and
all $i$.  Here $C_{\epsilon}(a_i)$ is the circle of radius $\epsilon$ about $
a_i$.

Next we introduce convenient orthornormal bases for
the subspaces that are of interest to us.
Define,
\[e_n^{(m)}(r,\Theta )=\left[\begin{array}{c}
e^{i(n-\frac 12)\Theta}\alpha_n^{(m)}(mr)\\
e^{i(n+\frac 12)\Theta}\beta_n^{(m)}(mr)\end{array}
\right],\]
where,
\begin{eqnarray*}
\alpha_n^{(m)}(mr)=\frac {I_{n-\frac 12}(mr)}{\sqrt {I_{n-\frac 1
2}^2(mr)+I_{n+\frac 12}^2(mr)}},\\
\beta_n^{(m)}(mr)=\frac {I_{n+\frac 12}(mr)}{\sqrt {I_{n-\frac 12}^
2(mr)+I_{n+\frac 12}^2(mr)}}.\end{eqnarray*}
Also define $e_n^{(m)*}(r,\Theta )=\left[\begin{array}{cc}
0&1\\
1&0\end{array}
\right]\bar {e}_n^{(m)}(r,\Theta )$.
The collection, 
\[\{e_n^{(m)}(\epsilon ,\Theta_j),e_n^{(m)*}(\epsilon ,\Theta_j)\}
,\]
as $n$ ranges over the integers is an orthonormal
basis for $L^2(C_{\epsilon}(a_j))$ (with values in ${\bf C}^2$).  Here we write, 
\[\Theta_j(z):=\Theta (z-a_j).\]
\begin{definition}\label{subspace1}
Let $W_j^{(m)}$ 
denote the subspace of $L^2(C_{\epsilon}(a_j))$ which is the $L^2$ closure
of the span of $e_n^{(m)}(\epsilon ,\Theta_j)$, and $e_n^{(m)*}(\epsilon 
,\Theta_j)$ for $n>0$  and the vector 
$e_0^{(m)}(\epsilon ,\Theta_j)+e_0^{(m)*}(\epsilon ,\Theta_j)$.  
\end{definition}
Define
\[\hat {e}_n^{(m)}(r,\theta )=\left[\begin{array}{c}
e^{-in\theta}\alpha_n^{\infty}(mr)\\
e^{-i(n-1)\theta}\beta_n^{\infty}(mr)\end{array}
\right],\]
with
\begin{eqnarray*}
\alpha_n^{\infty}(mr)=\frac {-K_n(mr)}{\sqrt {K_{n-1}^2(mr)+K_n^2
(mr)}},\\
\beta_n^{\infty}(mr)=\frac {K_{n-1}(mr)}{\sqrt {K_{n-1}^2(mr)+K_n^
2(mr)}}.\end{eqnarray*}
Then $\{\hat {e}_n^{(m)}(R,\theta )\}$, where $n$ ranges over the integers, is an
orthonormal set in $L^2(C_R).$ 
\begin{definition}\label{subspace2}
Let $W^{(m)}_{\infty}$ be the $L^2$ closure
of the span of the $\hat {e}_n^{(m)}(R,\theta )$ for $n\in {\bf Z}$.  
\end{definition}
The boundary 
conditions (2) and (3) in Lemma(\ref{pchar}) become,
\begin{eqnarray*}
\delta {\cal W}_j|_{C_{\epsilon}(a_i)}\in W^{(m)}_i,\\
\delta {\cal W}_j|_{C_R}\in W_{\infty}^{(m)}.\end{eqnarray*}
As a first step towards controlling the $m\rightarrow 0$ limit of the
solution of the boundary value problem described in 
Lemma(\ref{pchar}) we will now record some elementary
estimates for the convergence of $e_n^{(m)}(r,\Theta )$ and $\hat {
e}_n^{(m)}(r,\theta )$ to
their limits as $m\rightarrow 0$.  For $\ell >0$ the Bessel function $
I_{\ell}(r)$ 
behaves like a constant times $r^{\ell}$ as $r\rightarrow 0$. It immediately
follows that (for $n\ge 1$),  
\begin{eqnarray*}
\lim_{m\rightarrow 0}\alpha_n^{(m)}(mr)=1,\\
\lim_{m\rightarrow 0}\beta_n^{(m)}(mr)=0.\end{eqnarray*}
and hence that,
\[\lim_{m\rightarrow 0}e_n^{(m)}(r,\Theta )=e_n(\Theta ):=\left[\begin{array}{c}
e^{i(n-\frac 12)\Theta}\\
0\end{array}
\right]\mbox{\rm \ for $n\ge 1$.}\]
For similar reasons we find that,
\[\lim_{m\rightarrow 0}\hat {e}_n^{(m)}(r,\theta )=\hat {e}_n(\theta 
):=\left[\begin{array}{c}
-e^{-in\theta}\\
0\end{array}
\right]\mbox{\rm \ for }n\ge 1,\]
and recalling that $K_n(r)=K_{-n}(r)$ we find,
\[\lim_{m\rightarrow 0}\hat {e}_n^{(m)}(r,\theta )=\hat {e}_n(\theta 
):=\left[\begin{array}{c}
0\\
e^{-i(n-1)\theta}\end{array}
\right]\mbox{\rm \ for }n\le 0.\]

For $\ell >0$, $I_{\ell}(r)$ is an increasing function of $r$ and since
$I_{\ell}'(r)=-\frac {\ell}rI_{\ell}(r)+I_{\ell -1}(r)$ 
the right hand side is non negative and hence,
\begin{equation}\frac {I_{\ell}(r)}{I_{\ell -1}(r)}\le\frac r{\ell}\mbox{\rm \ for }
\ell >0.\label{1}\end{equation}
For $n>0$, $K_n(r)$ is a decreasing function of $r$ and since
$K_n'(r)=\frac nrK_n(r)-K_{n+1}(r)$ the right hand side is non 
positive and hence,
\begin{equation}\frac {K_n(r)}{K_{n+1}(r)}\le\frac rn\mbox{\rm \ for }
n>0.\label{2}\end{equation}
Now suppose that $0<a<b$, then we have,
\begin{equation}\frac a{\sqrt {a^2+b^2}}\le\frac ab,\label{3}\end{equation}
and
\begin{equation}1\ge\frac b{\sqrt {b^2+a^2}}=\frac 1{\sqrt {1+\left
(\frac ab\right)^2}}\ge 1-\frac 12\left(\frac ab\right)^2.\label{4}\end{equation}

Using equations (\ref{1}), (\ref{2}), (\ref{3}), and (\ref{4}),
we obtain the following estimates, 
\begin{equation}|\alpha_n^{(m)}(mr)-1|\le\frac 12\left(\frac {I_{
n+\frac 12}(mr)}{I_{n-\frac 12}(mr)}\right)^2\le\frac 12\left(\frac {
mr}{n+\frac 12}\right)^2\mbox{\rm \ for }n\ge 1,\label{5}\end{equation}
\begin{equation}|\beta_n^{(m)}(mr)|\le\frac {I_{n+\frac 12}(mr)}{
I_{n-\frac 12}(mr)}\le\frac {mr}{n+\frac 12}\mbox{\rm \ for }n\ge 
1,\label{6}\end{equation}
and
\begin{equation}|\alpha_n^{\infty}(mr)+1|\le\frac 12\left(\frac {
K_{n-1}(mr)}{K_n(mr)}\right)^2\le\frac 12\left(\frac {mr}{n-1}\right
)^2\mbox{\rm \ for }n\ge 2,\label{7}\end{equation}
\begin{equation}|\beta_n^{\infty}(mr)|\le\frac {K_{n-1}(mr)}{K_n(
mr)}\le\frac {mr}{n-1}\mbox{\rm \ for }n\ge 2.\label{8}\end{equation}
For $n\le -1$ we also have,
\begin{equation}|\alpha^{\infty}_n(mr)|\le\frac {K_{|n|}(mr)}{K_{
|n|+1}(mr)}\le\frac {mr}{|n|},\label{9}\end{equation}
and
\begin{equation}|\beta_n^{\infty}(mr)-1|\le\frac 12\left(\frac {K_{
|n|}(mr)}{K_{|n|+1}(mr)}\right)^2\le\frac 12\left(\frac {mr}{|n|}\right
)^2.\label{10}\end{equation}
We will use these bounds in the fifth section to 
estimate the norm of a graph operator for the inversion
of a certain projection.  The $n$ dependence in
these bounds will be of use to us there.  Note that 
estimates (\ref{7}) and (\ref{8}) obviously fail for $n=1.$
However, the first part of those estimates still hold for
$n=1$ and since,
\[\frac {K_0(mr)}{K_1(mr)}\stackrel {m\rightarrow 0}{\longrightarrow}
0,\]
it follows that $|\alpha_1^{\infty}(mr)+1|$ and $|\beta_1^{\infty}
(mr)|$ both tend to 0
as $m\rightarrow 0.$  For the same reason even though (\ref{9}) 
and(\ref{10}) are not valid for $n=0$ we still find that
$|\alpha_0^{\infty}(mr)|$ and $|\beta_0^{\infty}(mr)-1|$ both tend to 0 as $
m\rightarrow 0$.

We write
\[W^{(m)}:=W_{\infty}^{(m)}\oplus W_1^{(m)}\oplus\cdots\oplus W_N^{
(m)},\]
and $W^{(0)}$ for the $m\rightarrow 0$ limit of $W^{(m)}$.  The estimates
we've just given allow us to calculate the limiting
behavior of the basis elements of $W^{(m)}$.  This makes
it natural to define,
\begin{equation}W_j^{(0)}=\mbox{\rm span}_{n\ge 1}\left\{\left[\begin{array}{c}
e^{i(n-\frac 12)\Theta_j}\\
0\end{array}
\right],\left[\begin{array}{c}
0\\
e^{-i(n-\frac 12)\Theta_j}\end{array}
\right]\right\}\oplus {\bf C}\left[\begin{array}{c}
e^{-\frac {i\Theta_j}2}\\
e^{\frac {i\Theta_j}2}\end{array}
\right],\label{basis1}\end{equation}
for $j=1,\ldots ,N$ and
\begin{equation}W_{\infty}^{(0)}=\mbox{\rm span}_{n\ge 1}\left\{\left
[\begin{array}{c}
e^{-in\theta}\\
0\end{array}
\right]\right\}\oplus\mbox{\rm span}_{n\ge 1}\left\{\left[\begin{array}{c}
0\\
e^{in\theta}\end{array}
\right]\right\},\label{basis2}\end{equation}
with
\[W^{(0)}:=W_{\infty}^{(0)}\oplus W_1^{(0)}\oplus W_2^{(0)}\oplus
\cdots\oplus W_N^{(0)}.\]
To be a little more precise we will write $W^{(0)}$ for the
$L^2$ closure of the span of the basis vectors given above
(recall that $W^{(m)}$ was also a closed subspace of $L^2$).

We will always regard, 
\begin{eqnarray*}
W_j^{(0)}\subset L^2({\cal E}_{C_{\epsilon}(a_j)}),\\
W_{\infty}^{(0)}\subset L^2({\cal E}_{C_R}),\end{eqnarray*}
so that $W^{(0)}\subset L^2({\cal E}_{\partial\Omega})$.   Notice that the basis elements
of the subspaces $W_j^{(0)}$ can be regarded as {\em smooth }
sections of ${\cal E}_{C_{\epsilon}(a_j)}$ in the ${\cal U}_0$ trivialization.  Similarly
the basis elements of $W_{\infty}^{(0)}$ can be regarded as {\em smooth}
sections of ${\cal E}_{C_R}$ in the ${\cal U}_{\infty}$ trivialization.  We will do
this henceforth and it will make a difference for us
when we look at the subspaces of $W^{(0)}$ obtained by
taking the closure of the span of the same basis elements
in the Sobolev spaces $W^{\frac 12,p}({\cal E}_{\partial\Omega})$.

\section{The $m=0$ Green Function}

The Green function we want to understand has the
following matrix kernel,
\begin{equation}G_0(z,z')=-\frac 1{4\pi i}\left[\begin{array}{ccc}
\sum_ju_j(z)\overline {v_j(z')}&g(z,z')\\
\overline {g(z,z')}&\sum_j\overline {u_j(z)}v_j(z')&\end{array}
\right],\label{11}\end{equation}
where,
\begin{equation}u_j(z):=(z-a_j)^{-\frac 12}\prod_{k\ne j}\frac {(
z-a_k)^{\frac 12}}{(a_j-a_k)^{\frac 12}},\label{12}\end{equation}
\begin{equation}g(z,z'):=\sum_{|\epsilon |=0}c(\epsilon )\frac {\prod_
j(z-a_j)^{\epsilon_j}(z'-a_j)^{-\epsilon_j}}{z'-z},\label{13}\end{equation}
with $\epsilon =(\epsilon_1,\ldots ,\epsilon_n)$ and each $\epsilon_
j=\pm\frac 12$.  Also
\[|\epsilon |:=\sum_{j=1}^N\epsilon_j,\]
\begin{equation}c(\epsilon ):=\frac {\prod_{j<k}|a_j-a_k|^{2\epsilon_
j\epsilon_k}}{\sum_{|\epsilon |=0}\prod_{j<k}|a_j-a_k|^{2\epsilon_
j\epsilon_k}},\label{14}\end{equation}
and
\begin{equation}v_j(z)=(z-a_j)^{-\frac 12}\sum_{{|\epsilon 
|=0,\epsilon_j=\frac 12}{}}c(\epsilon )\prod_{k\ne j}\frac {(z-a_
k)^{\epsilon_k}}{(a_j-a_k)^{\epsilon_k}}.\label{15}\end{equation}
  The multivalued functions
$(z-a_j)^{\epsilon_j}$ are all defined using
the argument $\Theta_j$ and are consequently branched along 
$z\in {\bf r}_j$ .  We regard $G_0(z,z')$ as defining an operator, $
G_0,$
acting on sections of ${\cal E}_{\Omega}$ in the following manner,
\begin{equation}G_0f(z):=\int_{\Omega}G_0(z,z')f(z')dz'd\bar {z}'
,\label{16}\end{equation}
where the section $f(z')$ is identified with its ${\cal U}_0$ 
trivialization.  We also regard $G_0f$ as a section of
${\cal E}_{\Omega}$ given in the ${\cal U}_0$ trivialization. 

When working with $G_0f(z)$ for $|z|>R$ it is useful to 
rewrite $v_j(z)$ and $g(z,z')$ in terms appropriate for the
${\cal U}_{\infty}$ trivialization.  The conversion from the ${\cal U}_
0$ to
the ${\cal U}_{\infty}$ trivialization is given as follows,
\[\prod_k(z-a_k)^{\epsilon_k}\rightarrow\prod_k\left(1-\frac {a_k}
z\right)^{\epsilon_k}\mbox{\rm \ for }|\epsilon |=0\mbox{\rm \ and }
|z|>R,\]
where the fractional powers on the right are 
holomorphic functions functions of $z$ normalized to
be 1 at $z=\infty$.  In a similar fashion,
\begin{equation}v_j(z)\rightarrow z^{-1}\left(1-\frac {a_j}z\right
)^{-\frac 12}\sum_{|\epsilon |=0,\epsilon_j=\frac 12}c(\epsilon )
\prod_{k\ne j}\frac {\left(1-\frac {a_k}z\right)^{\epsilon_k}}{(a_
j-a_k)^{\epsilon_k}}\mbox{\rm \ for }|z|>R.\label{16a}\end{equation}

Our goal in this section is to show that $G_0$ inverts
$-\dsl$ with $W^{(0)}$ boundary conditions. It is precisely this 
property that determines our interest in $G_0$. 
We will at the same
time establish some elementary but useful estimates.

It is helpful to recall some well known results for the
kernel $\frac 1{z'-z}$.  Let $f\in C^1(\bar{\Omega })$ (the continuously differentiable 
functions on
$\Omega$ which are continuous together with their derivative 
on the closure of $\Omega$) and define,
\begin{equation}Tf(z):=\frac 1{2\pi i}\int_{\Omega}\frac {f(z')}{
z'-z}dz'd\bar {z}'\label{17}\end{equation}
Then 
\begin{theorem}\label{crinverse}
The distribution derivative of $Tf$ is
\begin{equation}dTf=T(\partial_zf)dz+fd\bar {z}.\label{18}\end{equation}
For $p>2$ one has the estimate,
\begin{equation}||Tf||_{W^{1,}{}^p(\Omega )}\le C_p||f||_{W^{1,p}
(\Omega )},\label{19}\end{equation}
for a constant $C_p$ that depends only on $p$ and $\Omega$, and $
W^{k,p}(\Omega )$ is
the subspace of $L^p(\Omega )$ which consists of functions whose 
first
$k$ weak derivatives are in $L^p(\Omega )$.
\end{theorem}
\emph{Proof}. Suppose $f\in C^1(\bar{\Omega })$ and $\phi\in C_0^{
\infty}(\Omega )$.  To
compute the distribution derivative $\partial_zTf(z)$ we wish
to ``integrate by parts'' in 
\[-\int_{\Omega}Tf(z)\partial_z\phi (z)\,dzd\bar {z}.\]

To stay away from the singularity in the kernel
we introduce,
\[T_{\epsilon}f(z):=\frac 1{2\pi i}\int_{\Omega\backslash D_{\epsilon}
(z)}\frac {f(z')}{z'-z}dz'd\bar {z}',\]
where $D_{\epsilon}(z)=\{z':|z'-z|<\epsilon \}$.  Also suppose for simplicity 
that $\epsilon$ is chosen small
enough so that the distance from the support of $\phi$ to
the boundary of $\Omega$ is greater than $\epsilon .$ For $\epsilon$ this small
it follows that for all $z$ in the support of $\phi$ the set
of $z'$ with $|z-z'|=\epsilon$ is completely contained in $\Omega 
.$

In order to do the ``integration by parts'' efficiently 
we calculate the exterior derivative of a particular 
3 form on the domain
$\Omega\times\Omega\backslash \{z=z'\},$
\begin{eqnarray*}
&&d\left(\frac {f(z')\phi (z)-f(z)\phi (z')}{z'-z}\right)d\bar {z}
dz'd\bar {z}'=\\
&&\left(\frac {f(z')\partial\phi (z)-\partial f(z)\phi (z')}{z'-z}
+\frac {f(z')\phi (z)-f(z)\phi (z')}{(z'-z)^2}\right)dzd\bar {z}d
z'd\bar {z}'.\end{eqnarray*}
Now integrate this last identity over,  
\[(\Omega\times\Omega )_{\epsilon}:=\Omega\times\Omega\backslash 
\{(z,z'):|z'-z|<\epsilon \},\]
and use Stokes' theorem.  Then make use of,
\[\int_{(\Omega\times\Omega )_{\epsilon}}\frac {f(z')\phi (z)-f(z
)\phi (z')}{(z'-z)^2}dzd\bar {z}dz'd\bar {z}'=0,\]
which follows from the fact that the integrand is odd 
under the transformation
$(z,z')\rightarrow (z',z)$ and the domain $(\Omega\times\Omega )_{
\epsilon}$ is invariant under
this map.  After multiplication by $\frac 1{2\pi i}$ the resulting identity
simplifies directly
to,
\begin{eqnarray*}
-\int_{\Omega}T_{\epsilon}f(z)\partial_z\phi (z)dzd\bar {z}=\int_{
\Omega}T_{\epsilon}(\partial f)(z)\phi (z)\,dzd\bar {z}\\
+\frac 1{2\pi i}\int_{\Omega}dz'd\bar {z}'\int_{|z-z'|=\epsilon}\frac {
f(z')\phi (z)-f(z)\phi (z')}{z'-z}d\bar {z}.\end{eqnarray*}
Since both $f$ and $\phi$ are $C^1(\Omega )$ it is easy to see that the
second term on the right hand side of this last equation
tends to 0 in the limit $\epsilon\rightarrow 0$.  Because $\frac 
1{z'-z}$ is in $L^1_{\mbox{\rm loc}}({\bf C}^2)$ 
it follows that $T_{\epsilon}f\rightarrow Tf$ in the sense of distributions as
$\epsilon\rightarrow 0.$ Hence,
\[\partial Tf=T\partial f,\]
which is the first part of (\ref{18}). To obtain the 
second part consider the exterior derivative,
\begin{eqnarray*}
d\left(\frac {f(z')\phi (z)}{z'-z}\right)dzdz'd\bar {z}'=-\bar{\partial}_
z\left(\frac {f(z')\phi (z)}{z'-z}\right)dzd\bar {z}dz'd\bar {z}'\\
=-\left(\frac {f(z')\bar{\partial}\phi (z)}{z'-z}\right)dzd\bar {
z}dz'd\bar {z}'.\end{eqnarray*}
Integrate this over $(\Omega\times\Omega )_{\epsilon}$ and multiply the result by 
$\frac 1{2\pi i}$.  Using Stokes' theorem again, one finds that,
\[-\int_{\Omega}T_{\epsilon}f(z)\bar{\partial}\phi (z)dzd\bar {z}
=-\frac 1{2\pi i}\int_{\Omega}dz'd\bar {z}'\int_{|z-z'|=\epsilon}\frac {
f(z')\phi (z)}{z'-z}dz.\]
As $\epsilon\rightarrow 0$ the right hand side tends to,
\[\int_{\Omega}dz'd\bar {z}'f(z')\phi (z').\]
Hence,
\[\bar{\partial }Tf=f,\]
which is the second part of (\ref{18}).

To prove the second part of the theorem we first 
observe that since $\Omega$ is a bounded domain it is
straightforward to show that
$T$ defines a bounded operator on $L^p(\Omega )$ for
$p>2$.  H\"older's inequality implies that,
\[\left|\int_{\Omega}\frac {f(z')}{z'-z}idz'd\bar z'\right|\le ||
f||_{L^p(\Omega )}\left(\int_{\Omega}|z'-z|^{-q}idz'd\bar z'\right
)^{\frac 1q},\]
where $\frac 1p+\frac 1q=1$.  However since $p>2$ it follows that
$1<q<2$ and hence that,
\[z\rightarrow\int_{\Omega}|z'-z|^{-q}idz'd\bar {z}',\]
is a bounded function on $\Omega$.  It follows at once that
$T$ is bounded on $L^p(\Omega )$ since $\Omega$ is a finite domain
and bounded functions are in $L^p(\Omega )$ (note:
the analogue of $T$ on $\Omega ={\bf C}$ is also bounded on $L^p(
{\bf C})$ for
$p>2$, (see \cite{IT92}).  

To see that it defines a bounded operator
on $W^{1,p}(\Omega )$ for $p>2$ it is enough to use (\ref{18}) which 
implies that for $f\in C^1(\bar{\Omega })$,
\[||dTf||_{L^p(\Omega )}\le ||T\partial f||_{L^p(\Omega )}+||f||_{
L^p(\Omega )}\le C||f||_{W^{1,p}(\Omega )}\mbox{\rm \ for }p>2\]
where we used the fact that $T$ is bounded on $L^p$ $(p>2)$.
Since the boundary of $\Omega$ is smooth, $C^1(\bar{\Omega })$ is dense in 
$W^{1,p}(\Omega )$, and the second part of the theorem is proved. 
QED

We will now use theorem (\ref{crinverse}) to establish
that $G_0$ is a Green function for $-\dsl$ on $W^{1,}{}^p({\cal E}_{
\Omega})$ for
$p>2$.  Incidentally, we work in the space $W^{1,p}({\cal E}_{\Omega}
)$ 
only in order to simplify some boundary estimates
by using the Sobolev trace theorems; we
could work in $L^p({\cal E}_{\Omega})$ at the cost of using more 
complicated global ellipticity estimates
(see \cite{BW93}). 
\begin{theorem}\label{greenfunction}
Suppose $p>2$ and suppose that $f\in W^{1,p}({\cal E}_{\Omega})$ and that
$f$ is compactly supported in $\Omega$.  Then $G_0f\in W^{1,p}({\cal E}_{
\Omega})$ and
\begin{list}{}{\setlength{\leftmargin}{.1in}}
\item{\bf 1.} $||G_0f||_{W^{1,p}}\le C_p||f||_{W^{1,p}}$
\item{\bf 2.} $-\dsl G_0f=f$
\item{\bf 3.} $G_0f|_{\partial\Omega}\in W^{(0)}$
\end{list}

\end{theorem}
\emph{Proof}.  Let $z_j^{\epsilon}=(z-a_j)^{\epsilon}$ be defined using the
argument $\Theta_j$ so that these functions are branched along
${\bf r}_j$.  For any choice $\epsilon_j=\pm\frac 12$ the function,
\[{\bf z}^{\epsilon}:=\prod_{j=1}^Nz_j^{\epsilon_j},\]
defines a map,
\[C^{\infty}({\cal E}_{\Omega})\ni f(z)\rightarrow {\bf z}^{\epsilon}
f(z)\in C^{\infty}(\Omega ),\]
which has an inverse,
\[C^{\infty}(\Omega )\ni f(z)\rightarrow {\bf z}^{-\epsilon}f(z)\in 
C^{\infty}({\cal E}_{\Omega}),\]
where in each case sections of $C^{\infty}({\cal E}_{\Omega})$ are identified
with their ${\cal U}_0$ trivializations.  Since the derivatives of
${\bf z}^{\pm\epsilon}$ are bounded on $\Omega$ it follows that these maps induce
bounded maps,
\[W^{1,p}({\cal E}_{\Omega})\ni f(z)\rightarrow {\bf z}^{\epsilon}
f(z)\in W^{1,p}(\Omega )\]
and
\[W^{1,p}(\Omega )\ni f(z)\rightarrow {\bf z}^{-\epsilon}f(z)\in 
W^{1,p}({\cal E}_{\Omega}).\]
The upper right matrix element of the kernel $G_0(z',z)$
is a linear combination of terms,
\[\frac {{\bf z}^{\epsilon}({\bf z}')^{-\epsilon}}{z'-z},\]
each of which is the kernel of an operator we can 
interpret as a composition,
\[W^{1,p}({\cal E}_{\Omega})\stackrel {{\bf z}^{-\epsilon}}{\longrightarrow}
W^{1,p}(\Omega )\stackrel {\frac 1{z'-z}}{\longrightarrow}W^{1,p}
(\Omega )\stackrel {{\bf z}^{\epsilon}}{\longrightarrow}W^{1,p}({\cal E}_{
\Omega}),\]
which is bounded for $p>2$ as a consequence of Theorem 
(\ref{crinverse}).  To be more precise we note that it
is the {\em line bundle} ${\cal E}_{\Omega}$ which appears in this composition.
The same argument for $\bar {{\bf z}}^{\pm\epsilon}$ and
$\frac 1{\bar {z}'-\bar {z}}$ coupled with the complex conjugate version of
Theorem (\ref{crinverse}) shows that the lower left
kernel in $G_0(z,z')$ determines a bounded linear 
transformation on $W^{1,p}({\cal E}_{\Omega})$.  The diagonal elements of
$G_0(z,z')$ are finite rank $L^2$ kernels with a range that consists
of smooth sections of ${\cal E}_{\Omega}.$  Consequently, they determine
bounded operators on $W^{1,p}({\cal E}_{\Omega})$ as well and this finishes
the proof of part (1) of the theorem.

The proof of part (2) of the theorem is a straightforward
computation using (1) of Theorem (\ref{crinverse}) (and
its complex conjugate), the fact that $\bar{\partial}_z{\bf z}^{\epsilon}
=0$, $\partial\bar {{\bf z}}^{\epsilon}=0$,
$\bar{\partial}_zu_j(z)=0$, $\partial_z\bar {u}_j(z)=0$, and finally, 
\[\sum_{|\epsilon |=0}c(\epsilon )=1.\]
Note that both $\bar{\partial}$ and $\partial$ act on $C^{\infty}
({\cal E}_{\Omega})$ since the 
transition functions that define the bundle are piecewise
constant.

To establish part (3) of the theorem it is useful to 
observe that the subspace $W^{(0)}_j$ consists of $L^2$ boundary
values on $C_{\epsilon}(a_j)$, of functions,
\[\left[\begin{array}{c}
z_j^{\frac 12}h_1(z)\\
\bar {z}^{\frac 12}_j\bar {h}_2(z)\end{array}
\right]+c_0\left[\begin{array}{c}
z_j^{-\frac 12}\\
\bar {z}_j^{-\frac 12}\end{array}
\right],\]
where $h_1$ and $h_2$ are holomorphic functions on the disk
$D_{\epsilon}(a_j)$ and $c_0$ is a complex 
constant (technically, $h_1$ and $h_2$ should be in the 
appropriate Hardy space).  We wish to show that
\[\int_{\Omega}G_0(z,z')f(z')dz'd\bar {z}',\]
restricted to $z\in C_{\epsilon}(a_j)$ lies in $W^{(0)}_j.$ Because
we have assumed that the support of $f$ is contained
inside $\Omega$ it follows that for $z'$ in the support of
$f$ we have $|z'-a_j|>|z-a_j|$ and so,
\[\frac 1{z'-z}=\frac 1{z'_j-z_j}=\sum_{n=0}^{\infty}\frac 1{z'_j}\left
(\frac {z_j}{z_j'}\right)^n,\]
will converge uniformly for $z\in D_{\epsilon}(a_j)$ and $z'$ in 
the support of $f$.  Substituting
this expansion in the formula for $G_0f(z)$ (and the
analogue obtained by taking complex conjugates) one
sees easily that the boundary value of $G_0f(z)$ has the
form,
\[\left[\begin{array}{c}
z_j^{-\frac 12}h_1(z)\\
\bar {z}^{-\frac 12}_j\bar {h}_2(z)\end{array}
\right],\]
where $h_1$ and $h_2$ are holomorphic in $D_{\epsilon}(a_j).$ The only 
issue is whether the coefficient of $z_j^{-\frac 12}$ in the first
component is the same as the coefficient of $\bar {z}_j^{-\frac 1
2}$ in
the second component.  A computation shows that
the coefficient of $z_j^{-\frac 12}$ in the Fourier expansion on $
C_{\epsilon}(a_j)$ of the 
first component of $G_0f$ is,
\[-\frac 1{4\pi i}\int_{\Omega}dz'd\bar {z}'\left\{\bar v_j(z')f_
1(z')+\sum_{{|\epsilon |=0,\epsilon_j=\frac 12}{}}c(\epsilon 
)z^{\prime -\frac 12}_j\prod_{k\ne j}\frac {(z'-a_k)^{\epsilon_k}}{
(a_j-a_k)^{\epsilon_k}}f_2(z')\right\}\]
where we used the fact that,
\[\sum_{|\epsilon |=0,\epsilon_j=-\frac 12}S(\epsilon )=\sum_{|\epsilon 
|=0,\epsilon_j=\frac 12}S(-\epsilon ).\]
Computing the coefficient of $\bar {z}_j^{-\frac 12}$ in the Fourier 
expansion of the second component of  $G_0f$ we find,
\[-\frac 1{4\pi i}\int_{\Omega}dz'd\bar {z}'\left\{v_j(z')f_2(z')
+\sum_{{|\epsilon |=0,\epsilon_j=\frac 12}{}}c(\epsilon 
)\bar z_j^{\prime -\frac 12}\prod_{k\ne j}\frac {\overline {(z'-a_
k)}\,^{\epsilon_k}}{\overline {(a_j-a_k)}\,^{\epsilon_k}}f_1(z')\right
\}.\]
Comparing these two coefficients using the definition of 
$v_j(z)$ we see that they are the same.  Thus 
$G_0f|_{C_{\epsilon}(a_j)}\in W^{(0)}_j$.  

To finish the proof we need to show that, 
\[G_0f|_{C_R}\in W^{(0)}_{\infty}.\]
Recalling the definition of $W_{\infty}^{(0)}$ we see that it consists
of boundary values on $C_R$ of functions,
\[\left[\begin{array}{c}
h_1(z)\\
\bar {h}_2(z)\end{array}
\right],\]
where $h_1(z)$ and $h_2(z)$ are holomorphic functions on $D_{\infty}$ which
vanish at $z=\infty$.  The condition $|\epsilon |=0$ in the sum that
defines $g(z,z')$ makes it easy to see that,
\[\int_{\Omega}g(z,z')f_2(z')dz'd\bar {z}',\]
is holomorphic for $z\in D_{\infty}$ and tends to 0 at $\infty$.  For
precisely the same reason,
\[\int_{\Omega}\overline {g(z,z')}f_1(z')dz'd\bar {z}',\]
is anti-holomorphic in $D_{\infty}$ and vanishes at $\infty .$  The 
diagonal contributions,
\[-\frac 1{4\pi i}\int_{\Omega}\sum_ju_j(z)\bar {v}_j(z')f_1(z')d
z'd\bar {z}',\]
and 
\[-\frac 1{4\pi i}\int_{\Omega}\sum_j\bar {u}_j(z)v_j(z')f_2(z')d
z'd\bar {z}',\]
do not at first appear to vanish at infinity since $u_j(z)$ 
does not tend to 0 at $\infty .$ However in the lemma which
follows this theorem we will prove the homogeneous
function identity,
\begin{equation}\sum_ju_j(z)\bar {v}_j(z')=\sum_jv_j(z)\bar {u}_j
(z').\label{20}\end{equation}
Since $v_j(z)$ is holomorphic for $|z|>R$ and does tend to 0 at $
\infty$ 
(see (\ref{16a})) this identity finishes the
proof of the theorem. QED

We turn to the proof of the identity (\ref{20}). 
\begin{lemma}\label{homo} The following identity is true
for the functions $u_j$ and $v_j$ defined in (\ref{12}) and (\ref{15})
above,
\[\sum_ju_j(z)\bar {v}_j(z')=\sum_jv_j(z)\bar {u}_j(z').\]
\end{lemma}
\emph{Proof}. Suppose that $v(z)$ is a holomorphic function
branched along the rays ${\bf r}_j$ such that,
\[V(z)=v(z)\prod_kz_k^{-\frac 12},\]
is holomorphic in the punctured plane, ${\bf C}\backslash {\bf a}$, with simple
poles at each $a_j$ and which tends to 0 as $z\rightarrow\infty$ (this
will be the case for each of the functions $v_j$).  Then
$V(z)$ has the partial fraction decomposition,
\[V(z)=\sum_k\frac {V_k}{z-a_k},\]
where $V_k=\mbox{\rm Res}_{z=a_k}V(z)$.  Rewriting this in terms of 
$v(z)$ one finds,
\[v(z)=\sum_kV_kz_k^{-\frac 12}\prod_{\ell\ne k}z_{\ell}^{\frac 1
2}=\sum_kV_k\prod_{\ell\ne k}(a_k-a_{\ell})^{\frac 12}u_k(z).\]
Thus we have,
\[v_j(z)=\sum_kv_{kj}u_k(z),\]
where the coefficients $v_{kj}$ are found by residue 
calculation to be,
\[v_{jj}=\sum_{{|\epsilon |=0,\epsilon_j=\frac 12}}c(
\epsilon ),\]
and for $k\ne j$, 
\begin{equation}v_{kj}=\frac {(a_k-a_j)^{\frac 12}}{(a_j-a_k)^{\frac 
12}}\sum_{{{|\epsilon |=0}\atop {\epsilon_j=\frac 12,\epsilon_k=-\frac 
12}}}c(\epsilon )\prod_{\ell\ne j,k}\frac {(a_k-a_{\ell})^{\epsilon_{
\ell}}}{(a_j-a_{\ell})^{\epsilon_{\ell}}}.\label{vres}\end{equation}
We will now show that $v_{kj}=\bar {v}_{jk}$ which will have 
(\ref{20}) as a simple consequence.  Since $v_{jj}$ is real
it is clear that $v_{jj}=\bar {v}_{jj}$.  Now suppose that $k\ne 
j$ and
 note that,
\[\frac {(a_k-a_j)^{\frac 12}}{(a_j-a_k)^{\frac 12}}=\frac {\overline {
(a_j-a_k)^{\frac 12}}}{\overline {(a_k-a_j)^{\frac 12}}},\]
since cross multiplication produces the identity,
\[|a_k-a_j|=|a_j-a_k|.\]
Thus the first factor in (\ref{vres}) is Hermitian 
symmetric and it remains only to check that the second
factor is also.  First we rewrite $c(\epsilon )$ for $\epsilon_j=\frac 
12$ and 
$\epsilon_k=-\frac 12$ in the following manner, 
\[c(\epsilon )_{{{\epsilon_j=\frac 12}\atop {\epsilon_k=-\frac 12}}}
=\frac {|a_j-a_k|^{-\frac 12}}{{\bf c}}\prod_{{{\alpha <\beta}\atop {
\alpha ,\beta\notin \{j,k\}}}}|a_{\alpha}-a_{\beta}|^{2\epsilon_{
\alpha}\epsilon_{\beta}}\prod_{\ell\neq j,k}\frac {|a_j-a_{\ell}|^{
\epsilon_{\ell}}}{|a_k-a_{\ell}|^{\epsilon_{\ell}}},\]
where ${\bf c}:=\sum_{\alpha <\beta}|a_{\alpha}-a_{\beta}|^{2\epsilon_{
\alpha}\epsilon_{\beta}}$.  Now define,
\[s_{kj}(\epsilon )=\frac {|a_j-a_k|^{-\frac 12}}{{\bf c}}\prod_{{{
\alpha <\beta}\atop {\alpha ,\beta\notin \{j,k\}}}}|a_{\alpha}-a_{
\beta}|^{2\epsilon_{\alpha}\epsilon_{\beta}}.\]
The second factor in (\ref{vres}) becomes,
\[\sum_{{{|\epsilon |=0}\atop {\epsilon_j=\frac 12,\epsilon_k=-\frac 
12}}}s_{kj}(\epsilon )\prod_{\ell\ne j,k}\frac {|a_j-a_{\ell}|^{\epsilon_{
\ell}}}{|a_k-a_{\ell}|^{\epsilon_{\ell}}}\frac {(a_k-a_{\ell})^{\epsilon_{
\ell}}}{(a_j-a_{\ell})^{\epsilon_{\ell}}}.\]
Since $s_{kj}(\epsilon )$ is real and obviously equal to $s_{jk}(
\epsilon )$ this
last expression will be Hermitian symmetric provided
that,
\[\prod_{\ell\ne j,k}\frac {|a_j-a_{\ell}|^{\epsilon_{\ell}}}{|a_
k-a_{\ell}|^{\epsilon_{\ell}}}\frac {(a_k-a_{\ell})^{\epsilon_{\ell}}}{
(a_j-a_{\ell})^{\epsilon_{\ell}}}=\prod_{\ell\ne j,k}\frac {|a_k-
a_{\ell}|^{\epsilon_{\ell}}}{|a_j-a_{\ell}|^{\epsilon_{\ell}}}\frac {\overline {
(a_j-a_{\ell})^{\epsilon_{\ell}}}}{\overline {(a_k-a_{\ell})^{\epsilon_{
\ell}}}}.\]
But this follows directly from cross multiplication as
before.  We've shown that $v_{kj}=\bar {v}_{jk}$ and the following
simple calculation now proves the identity (\ref{20}),
\begin{eqnarray*}
\sum_ju_j(z)\bar {v}_j(z')=\sum_{j,k}u_j(z)\bar {v}_{kj}\bar {u}_
k(z')\\
=\sum_{j,k}v_{jk}u_j(z)\bar {u}_k(z')=\sum_kv_k(z)\bar {u}_k(z').\end{eqnarray*}
QED
\section{The Projection on $W^{(0)}$}
In this section we will introduce the projection $P_0$ from
$L^2({\cal E}_{\partial\Omega})$ onto $W^{(0)}$ which is naturally associated with the
Green function $G_0$.  Another goal is a description of the 
complementary subspace for $P_0$ acting on $H^{\frac 12}({\cal E}_{
\partial\Omega}).$ We
will show that the complementary projection $I-P_0$ 
projects $H^{\frac 12}({\cal E}_{\partial\Omega})$ onto the boundary values of sections
$\Psi\in H^1({\cal E}_{\Omega})$ which are solutions to the Dirac equation,
$\dsl\Psi =0$ in $\Omega$.

It is useful to start with a calculation.  
Write,
\[G_0(z,z')=\left[\begin{array}{cc}
G_{11}(z,z')&G_{12}(z,z')\\
G_{21}(z,z')&G_{22}(z,z')\end{array}
\right],\]
for the matrix elements of $G_0$.

Now suppose that $f\in C^1({\cal E}_{\bar{\Omega}})$ and choose $
z\in {\bf C}\backslash\bar{\Omega}$.  Then, 
\begin{eqnarray*}
-G_0(\dsl f)(z)&=&-2\int_{\Omega}\left[\begin{array}{cc}
G_{11}(z,z')\partial_{z'}f_2(z')+G_{12}(z,z')\bar{\partial}_{z'}f_
1(z')&\\
G_{21}(z,z')\partial_{z'}f_2(z')+G_{22}(z,z')\bar{\partial}_{z'}f_
1(z')&\end{array}
\right]dz'd\bar {z}'\\
&=&-2\int_{\Omega}d_{z'}\left[\begin{array}{c}
G_{11}(z,z')f_2(z')d\bar {z}'-G_{12}(z,z')f_1(z')dz'\\
G_{21}(z,z')f_2(z')d\bar {z}'-G_{22}(z,z')f_1(z')dz'\end{array}
\right],\end{eqnarray*}
which Stokes' theorem transforms into,
\[-G_0(\dsl f)(z)=-2\int_{\partial\Omega}\left[\begin{array}{c}
G_{11}(z,z')f_2(z')d\bar {z}'-G_{12}(z,z')f_1(z')dz'\\
G_{21}(z,z')f_2(z')d\bar {z}'-G_{22}(z,z')f_1(z')dz'\end{array}
\right].\]
The first equality follows from the fact that $G_{k2}(z,z')$ 
is holomorphic for $z'\in\Omega$ and $G_{k1}(z,z')$ is anti-holomorphic in 
$z'\in\Omega$ (remember $z$ is outside of $\Omega$). Of course, this is not 
precisely accurate since these
functions are branched along the rays ${\bf r}_j$.  However, it is
not hard to argue that the application of Stokes' 
theorem is still correct using the fact that $G_{k2}(z,z')f_1(z')$
and $G_{k1}(z,z')f_2(z')$ are continuous for $z'$ on the rays ${\bf r}_
j$.
Also note that the orientation of $\partial\Omega$ appropriate for this
Stokes' calculation is that $C_R$ has the usual 
counterclockwise orientation and the circles $C_{\epsilon}(a_j)$ are all
{\em clockwise oriented.\/}  With this as our motivation, we 
define, for $f\in L^2({\cal E}_{\partial\Omega}),$
\begin{equation}P_0f(z):=-2\int_{\partial\Omega}\left[\begin{array}{c}
G_{11}(z,z')f_2(z')d\bar {z}'-G_{12}(z,z')f_1(z')dz'\\
G_{21}(z,z')f_2(z')d\bar {z}'-G_{22}(z,z')f_1(z')dz'\end{array}
\right].\label{proj}\end{equation}
where we understand this as a section of ${\cal E}_{\partial\Omega}$ by letting
$z\rightarrow\partial\Omega$ from outside of $\Omega$.  As usual sections of $
{\cal E}$ are
identified with their ${\cal U}_0$ or ${\cal U}_{\infty}$ trivializations. If $
-\dsl f$ is
compactly supported in $\Omega$ we saw in the last section 
that the restriction of $-G_0\dsl f$ to $\partial\Omega$ is in $W^{
(0)}$.  Thus
(\ref{proj}) suggests that if $f|_{\partial\Omega}\in W^{(0)}$ we should have
$P_0f=f.$ Our first result in this section is,
\begin{theorem}\label{P0}
The map $P_0$ defined by (\ref{proj}) is a projection 
from $L^2({\cal E}_{\partial\Omega})$ onto $W^{(0)}$.  $P_0$ restricts to a continuous
map,
\[P_0:H^{\frac 12}({\cal E}_{\partial\Omega})\rightarrow H^{\frac 
12}({\cal E}_{\partial\Omega}).\]
\end{theorem}
\emph{Proof}.  We will show that $P_0$ maps 
$L^2({\cal E}_{\partial\Omega})$ into $W^{(0)}$.  Observe first that the functions, $
f,$
in $L^2({\cal E}_{C_{\epsilon}(a_j)})$ which are restrictions to $
z\in C_{\epsilon}(a_j)$ of
the type
\begin{equation}f(z)=\sum_{n=-L}^L\left[\begin{array}{c}
f_{1n}z^{n+\frac 12}_j\\
f_{2n}\bar {z}^{n+\frac 12}_j\end{array}
\right],\label{21}\end{equation}
for $L$ finite,
are dense in $L^2({\cal E}_{C_{\epsilon}(a_j)})$, and have extensions to 
${\bf C}\backslash {\bf r}_j$ which are solutions to $\dsl f=0$. If $
\varphi$ is a $C_0^{\infty}$ function which
is 1 for $|z|\le 1.5\epsilon$ and 0 for $|z|>2\epsilon$ and $f$ is a function of 
type (\ref{21}) then it is easy to see that 
$\varphi_jf(z):=\varphi (z_j)f(z)$  
is a 
section of ${\cal E}_{\Omega}$ and $\dsl(\varphi_jf)$ is compactly supported inside
 $\Omega$.  For such a function the calculation that we began 
this section with shows that,
\[-G_0\dsl(\varphi_jf)|_{\partial\Omega}=P_0f,\]
and it follows from Theorem (\ref{greenfunction}) that
 $P_0f\in W^{(0)}.$  The first part of the theorem now follows 
from the fact that $W^{(0)}$ is closed in $L^2({\cal E}_{\partial
\Omega})$ and $P_0$ is
continuous on $L^2$.  We won't bother to give the proof 
that $P_0$ is continuous on $L^2$ since the argument we now
present to show that $P_0$ is continuous on $H^{^{}\frac 12}({\cal E}_{
\partial\Omega})$ adapts
directly to show $L^2$ continuity.  

Observe first that the finite rank part of $P_0$ associated
with the kernels $G_{11}$ and $G_{22}$ has a range which is a 
subset of $C^{\infty}({\cal E}_{\partial\Omega})\subset H^{\frac 
12}({\cal E}_{\partial\Omega})$ and is clearly continuous in
$L^2$ and hence also in $H^{\frac 12}$.  Next consider the part of $
P_0$ 
associated with $G_{12}$.  The component of this operator
which maps $H^{\frac 12}({\cal E}_{C_{\epsilon}(a_j)})$ into $H^{\frac 
12}$$({\cal E}_{C_{\epsilon}(a_k)})$ can be written
as a sum of operators each of which has a factorization
of the following sort, 
\begin{eqnarray*}
H^{\frac 12}({\cal E}_{C_{\epsilon}(a_j)})\stackrel {z_j^{\frac 1
2}}{\longrightarrow}H^{\frac 12}(C_{\epsilon}(a_j))\stackrel {\phi 
(z)}{\longrightarrow}H^{\frac 12}(C_{\epsilon}(a_j))\\
\stackrel {\frac 1{z'-z}}{\longrightarrow}H^{\frac 12}(C_{\epsilon}
(a_k))\stackrel {\psi (z)}{\longrightarrow}H^{\frac 12}(C_{\epsilon}
(a_k))\stackrel {z_k^{\frac 12}}{\longrightarrow}H^{\frac 12}({\cal E}_{
C_{\epsilon}(a_k))}),\end{eqnarray*}
where the first, second, fourth and fifth maps are
multiplication operators and the third map is,
\begin{equation}f(z)\rightarrow\frac 1{2\pi i}\int_{C_{\epsilon}(
a_j)}\frac {f(z')}{z'-z}\,dz',\label{22}\end{equation}
which must be interpreted as a suitable boundary value
when $j=k$.  In this factorization $\phi (z)$ and $\psi (z)$ are 
smooth functions and hence determine bounded maps on
$H^{\frac 12}$.  The Cauchy projection (\ref{22}) is easily seen to
be continuous from $H^{\frac 12}(C_{\epsilon}(a_j))$ to $H^{\frac 
12}(C_{\epsilon}(a_k))$ even when
$j=k$.  Nothing changes if $C_R$ is
one or both of the two components of $\partial\Omega$ that are 
involved and it follows
that the part of $P_0$ associated with $G_{12}$ is bounded
on $H^{\frac 12}({\cal E}_{\partial\Omega})$.  The kernel $G_{21}$ is just the complex
conjugate of $G_{12}$ and so it too defines a bounded
operator on $H^{\frac 12}({\cal E}_{\partial\Omega})$.  This completes the proof that $
P_0$ 
is continuous on $H^{\frac 12}({\cal E}_{\partial\Omega}).$

To finish the proof of the theorem we need to show 
that if $f\in W^{(0)}$ then $P_0f=f.$ Clearly it is enough to
show this for the basis elements (\ref{basis1}) and 
(\ref{basis2}).  For the calculation on $W^{(0)}_j$ it is preferable to
use the alternate forms for $G_{11}$ and $G_{22}$ found in 
Lemma(\ref{homo}).  Thus (\ref{proj}) becomes,
\begin{equation}P_0f(z)=\frac 1{2\pi i}\int_{\partial\Omega}\left
[\begin{array}{c}
\sum_kv_k(z)\bar {u}_k(z')f_2(z')d\bar {z}'-g(z,z')f_1(z')dz'\\
\overline {g(z,z')}f_2(z')d\bar {z}'-\sum_k\bar {v}_k(z)u_k(z')f_
1(z')dz'\end{array}
\right].\label{proj1}\end{equation}
The following residue calculations suffice to evaluate $P_0$ on the
basis elements of $W^{(0)}_j$ (note that in these results $C_{\epsilon}
(a_j)$ 
is counterclockwise oriented, as usual),
\[\frac 1{2\pi i}\int_{C_{\epsilon}(a_j)}u_k(z)z_j^{n-\frac 12}dz
=\delta_{jk}\delta_{n0}\mbox{\rm \ for $n=0,1,2,\ldots$}\]
And for $n=0,1,2,\ldots ,$
\begin{eqnarray*}
\frac 1{2\pi i}\int_{C_{\epsilon}(a_j)}\sum_{|\epsilon |=0}c(\epsilon 
)\frac {\prod_kz_k^{\epsilon_k}(z'_k)^{-\epsilon_k}}{z'-z}(z')^{n
-\frac 12}dz'\\
=\left\{\begin{array}{cc}
z_j^{n-\frac 12}-\delta_{n0}v_j(z)&\mbox{\rm for }z\in C_{\epsilon}
(a_j)\\
-\delta_{n0}v_j(z)&\mbox{\rm for }z\in\partial\Omega\backslash C_{
\epsilon}(a_j)\end{array}
\right.\end{eqnarray*}
One finds (being careful to recall the orientation of the
$C_{\epsilon}(a_j)$ component of $\partial\Omega$ is clockwise) that $
P_0$ fixes the
elements of the basis for $W^{(0)}_j$.  The reader might find
the cancelation of the $v_j(z)$ terms that appear in the
calculation of the action of $P_0$ on $\left[\begin{array}{c}
z_j^{-\frac 12}\\
\bar {z}_j^{-\frac 12}\end{array}
\right]$ instructive.

To compute the action of $P_0$ on the basis elements for
$W^{(0)}_{\infty}$ the original form for the kernel of the Green 
function $G_0$ is preferable and one can do the needed
calculation with the following residues,
\[\frac 1{2\pi i}\int_{C_R}v_k(z)z^{-n}dz=0\mbox{\rm \ for }n=1,2
,3,\ldots\]
which follows from the fact that $v_k(z)$ is holomorphic in
the exterior of $C_R$ and vanishes at $\infty$ in the ${\cal U}_{
\infty}$ trivialization.  
And for $n=1,2,3,\ldots ,$
\[\frac 1{2\pi i}\int_{C_R}\sum_{|\epsilon |=0}c(\epsilon )\frac {
\prod_kz_k^{\epsilon_k}(z'_k)^{-\epsilon_k}}{z'-z}(z')^{-n}dz'=\left
\{\begin{array}{c}
z^{-n}\mbox{\rm \ for }z\in C_R\\
0\mbox{\rm \ for }z\in\partial\Omega\backslash C_R\end{array}
\right.\]
Again one finds that $P_0$ fixes the basis (\ref{basis2}) and
this finishes the proof of the theorem. QED

Next we turn to a characterization of the 
complementary projection $I-P_0.$ 
\begin{theorem}\label{complement}
The projection $I-P_0$ maps $H^{\frac 12}({\cal E}_{\partial\Omega}
)$ into the subspace
of $H^{\frac 12}({\cal E}_{\partial\Omega})$ which consists of boundary values of 
functions $\Psi\in H^1({\cal E}_{\Omega})$ which satisfy the Dirac equation
\[\dsl\Psi =0\]
in $\Omega$.  Furthermore, there exists a constant $C$ so that 
for all $f\in H^{\frac 12}({\cal E}_{\partial\Omega})$ we have,
\begin{equation}||(I-P_0)f||_{H^1({\cal E}_{\Omega})}\le C||f||_{
H^{\frac 12}({\cal E}_{\partial\Omega})}\label{23}\end{equation}
\end{theorem}
\emph{Proof}.  Using (\ref{proj1}) and the well known
identity,
\[(z'-z_{int})^{-1}-(z'-z_{ext})^{-1}=2\pi i\delta (z'-z),\]
for the difference of the interior and exterior boundary
values of the Cauchy kernel on a circle we find the following 
formula for $P_0^c:=I-P_0,$
\begin{equation}P^c_0f(z)=-\frac 1{2\pi i}\int_{\partial\Omega}\left
[\begin{array}{c}
\sum_kv_k(z)\bar {u}_k(z')f_2(z')d\bar {z}'-g(z,z')f_1(z')dz'\\
\overline {g(z,z')}f_2(z')d\bar {z}'-\sum_k\bar {v}_k(z)u_k(z')f_
1(z')dz'\end{array}
\right],\label{proj2}\end{equation}
with the difference compared to (\ref{proj1}) that 
in this formula $z$ is to tend to
$\partial\Omega$ from the interior of $\Omega$.  It is clear from this formula
that $(I-P_0)f(z)$ extends to a section of ${\cal E}_{\Omega}$ which is in 
the null space of $\dsl$.  We need only establish the 
estimate (\ref{23}) to finish the proof.  The finite
rank operator,
\[f\rightarrow -\frac 1{2\pi i}\int_{\partial\Omega}\sum_kv_k(z)\bar {
u}_k(z')f(z')d\bar {z}',\]
is obviously continuous on $H^{\frac 12}({\cal E}_{\partial\Omega}
)$ since each $u_k$ is
in $L^2$ and the functions $v_k\in C^{\infty}({\cal E}_{\Omega})\subset 
H^1({\cal E}_{\Omega}).$  The
other finite rank operator that occurs in (\ref{proj2}) is
continuous from $H^{\frac 12}({\cal E}_{\partial\Omega})$ into $H^
1({\cal E}_{\Omega})$ for the same reason.
Next we turn to the operator,
\[f\rightarrow\frac 1{2\pi i}\int_{\partial\Omega}g(z,z')f(z')dz'
.\]
This operator is linear combination of operators each of which we
wish to interpret as a composition,
\[H^{\frac 12}({\cal E}_{\partial\Omega})\stackrel {{\bf z}^{-\epsilon}}{
\longrightarrow}H^{\frac 12}(\partial\Omega )\stackrel {\frac 1{z'
-z}}{\longrightarrow}H^1(\Omega )\stackrel {{\bf z}^{\epsilon}}{\longrightarrow}
H^1({\cal E}_{\Omega}).\]
The first and third maps are multiplication operators 
which are obviously continuous.  The middle map is
shorthand for the operator,
\[f\rightarrow\frac 1{2\pi i}\int_{\partial\Omega}\frac {f(z')}{z'
-z}dz',\]
which is well known to be continuous from $H^{\frac 12}(\partial\Omega 
)$ to
$H^1(\Omega ).$  For the reader's convenience we recall a simple
argument for this continuity.

Write $z_j=z-a_j$ and consider a function, $f$, defined
on $\partial\Omega$ by, 
\begin{equation}f(z)=\left\{\begin{array}{c}
\sum_{n=-L}^Lf_nz_j^n\mbox{\rm \ for }z\in C_{\epsilon}(a_j)\\
0\mbox{\rm \ for }z\in\partial\Omega\backslash C_{\epsilon}(a_j)\end{array}
\right.\label{24}\end{equation}

 Define $P_{int}f(z)$ for $z$ in $\Omega$ by, 

\[P_{int}f(z)=\frac 1{2\pi i}\int_{\Omega}\frac {f(z')}{z'-z}dz'.\]
Then 
\[P_{int}f(z)=\sum_{n=-L}^{-1}f_nz^n_j.\]
Since $\Omega$ is bounded the Poincare' inequality \cite{LL97} implies that the $
H^1(\Omega )$ 
norm of $P_{int}f$ is bounded by 
the $L^2(\Omega )$ norm of,
\[\partial P_{int}f(z)=\sum_{n=-L}^{-1}nf_nz^{n-1}_j.\]
The $L^2(\Omega )$ norm of $\partial P_{int}f$ is in turn dominated by the
$L^2$ norm of $\partial P_{int}f$ on $|z_j|\ge\epsilon$.  To compute this norm
it suffices to observe that (for $n,m=-1,-2,\ldots$),
\begin{eqnarray*}
\int_{|z_j|\ge\epsilon}z^{n-1}_j\bar {z}^{m-1}_jidzd\bar {z}=\int_{
|z_j|\ge\epsilon}d\left(\frac {z^n_j\bar {z}^{m-1}_j}nid\bar z\right
)\\
=-\frac 1n\int_{C_{\epsilon}(a_j)}z_j^n\bar {z}_j^{m-1}id\bar {z}
=\frac {2\pi}{|n|}\epsilon^{2n}\delta_{nm}.\end{eqnarray*}
We see then that,
\[||P_{int}f||_{H^1(\Omega )}^2\le C\sum_{n=-L}^{-1}\epsilon^{2n}
|n|f_n\bar {f}_n\le C||f||^2_{H^{\frac 12}(C_{\epsilon}(a_j))}.\]
Now suppose that,
\begin{equation}f(z)=\left\{\begin{array}{c}
\sum_{n=-L}^Lf_nz^n\mbox{\rm \ for }z\in C_R\\
0\mbox{\rm \ for }z\in\partial\Omega\backslash C_R.\end{array}
\right.\label{25}\end{equation}
Then the argument we just gave is easily modified to 
show that,
\[||P_{int}f||_{H^1(\Omega )}\le C||f||_{H^{\frac 12}(C_R)}.\]
Sums of functions of type (\ref{24}) for $j=1,2,\ldots ,N$ and
type (\ref{25}) are dense in $H^{\frac 12}(\partial\Omega )$ and it follows that
$P_{int}$ extends to a continuous map from $H^{\frac 12}(\partial
\Omega )$ to $H^1(\Omega )$.

Taking complex conjugates the result we just proved
also shows that the map,
\[f\rightarrow -\frac 1{2\pi i}\int_{\partial\Omega}\overline {g(
z,z')}f(z')d\bar {z}',\]
is bounded from $H^{\frac 12}({\cal E}_{\partial\Omega})$ to $H^1
({\cal E}_{\Omega})$.  This finishes the 
proof of the theorem. QED

\section{Inverting the Projection $P_0:W^{(m)}\rightarrow W^{(0)}$}
\emph{Remark}. In this section we will write $W^{(m)}$ for
$W^{(m)}\cap H^{\frac 12}({\cal E}_{\partial\Omega})$ and $W^{(0)}$ for $
W^{(0)}\cap H^{\frac 12}({\cal E}_{\partial\Omega})$.  We will
prove the following theorem. 

\begin{theorem}\label{isoproj}
For all sufficiently small values of $m$ the projection
\[P_0:W^{(m)}\rightarrow W^{(0)},\]
is an isomorphism.  Furthermore, there is a
linear map $\delta$ from $W^{(0)}$ into $(I-P_0)H^{\frac 12}({\cal E}_{
\partial\Omega})$
and a constant $C$ which is independent of $f$ and $m$ 
so that for all $f\in W^{(0)}$,

\begin{list}{}{\setlength{\leftmargin}{.1in}}
\item{\bf 1.} $f+\delta f\in W^{(m)}$
\item{\bf 2.} $||\delta f||_{H^{\frac 12}({\cal E}_{\partial\Omega}
)}\le Cm||f||_{H^{\frac 12}({\cal E}_{\partial\Omega})}$
\end{list}

\end{theorem}

\emph{Remark} The map $\delta$ gives $W^{(m)}$ as a graph over
$W^{(0)}.$

\emph{Proof}.  
Suppose first that $f\in W^{(m)}$ 
and let $f_j$ denote 
the restriction of $f$ to $C_{\epsilon}(a_j)$ and let $f_{\infty}$ denote the
restriction of $f$ to $C_R.$ Write $e_{n,j}^{(m)}$ for $e^{(m)}_n
(\epsilon ,\Theta_j)$ and
$\hat {e}^{(m)}_{n,\infty}$ for $\hat {e}_n^{(m)}(R,\theta )$.
Then the Fourier expansions of $f$ on $C_{\epsilon}(a_j)$ and $C_
R$ can 
be written,
\begin{equation}f_j=a_{0,j}(e_{0,j}^{(m)}+e_{0,j}^{(m)*})+\sum_{n
=1}^{\infty}\left\{a_{n,j}e^{(m)}_{n,j}+b_{n,j}e^{(m)*}_{n,j}\right
\}\label{26}\end{equation}
and
\begin{equation}f_{\infty}=\sum_{n=-\infty}^{\infty}a_{n,\infty}\hat {
e}^{(m)}_{n,\infty}\label{27}\end{equation}
Now let $p_0f$ denote the element of $W^{(0)}$ which has the 
same ``Fourier coefficients'' in the $m\rightarrow 0$ limiting basis.  
That is,
\begin{equation}p_0f_j=a_{0,j}(e_{0,j}+e_{0,j}^{*})+\sum_{n=1}^{\infty}\left
\{a_{n,j}e_{n,j}+b_{n,j}e^{*}_{n,j}\right\}\label{28}\end{equation}
and 
\begin{equation}p_0f_{\infty}=\sum_{n=-\infty}^{\infty}a_{n,\infty}
\hat {e}_{n,\infty},\label{29}\end{equation}
where $e_{n,j}$ denotes the basis vector $e_n(\Theta_j)$ and $\hat {
e}_{n,\infty}$ 
denotes the vector $\hat {e}_n(\theta )$.  It is easy to check that,  
\[<e_{k,j}^{(m)}-e_{k,j},e_{\ell ,j}^{(m)}-e_{\ell ,j}>_{H^{\frac 
12}({\cal E}_{\partial\Omega})}=\delta_{k\ell}||e_{k,j}^{(m)}-e_{
k,j}||^2_{H^{\frac 12}({\cal E}_{\partial\Omega})},\]
and from (\ref{5}) and (\ref{6}) it follows that,
\[||e_{k,j}^{(m)}-e_{k,j}|_{}|_{H^{\frac 12}({\cal E}_{\partial\Omega}
)}\le Cm\]
Analogous results for $e^{(m)*}_{k,j}$ and for $\hat {e}_k^{(m)}$ (which follow
from (\ref{7}) and (\ref{8})) imply the inequality,
\begin{equation}||p_0f-f||_{H^{\frac 12}({\cal E}_{\partial\Omega}
)}\le Cm||f||_{H^{\frac 12}({\cal E}_{\partial\Omega})},\label{30}\end{equation}
for some constant $C$ and all $f\in W^{(m)}.$   Now write
$\Delta p_0=p_0-I$.  Then for $f\in W^{(m)}$ we have,
\[f+\Delta p_0f\in W^{(0)}.\]
It follows from this that,
\[(I-P_0)(f+\Delta p_0f)=0,\]
or
\[(I-P_0)f=-(I-P_0)\Delta p_0f.\]
From this (\ref{30}) and the fact that $P_0$ is bounded on
$H^{\frac 12}({\cal E}_{\partial\Omega})$ it follows that, 
\[||f-P_0f||_{H^{\frac 12}({\cal E}_{\partial\Omega})}\le Cm||f||_{
H^{\frac 12}({\cal E}_{\partial\Omega})},\]
for some constant $C$ and all $f\in W^{(m)}.$  When $m$ is
small enough so that $Cm<1$ this last inequality implies
that $P_0:W^{(m)}\rightarrow W^{(0)}$ is injective since $P_0f=0$ gives
$||f||\le Cm||f||$ with $Cm<1$ which in turn forces $||f||=0$. 

Now start with $f_0\in W^{(0)}$ with Fourier
expansion given by (\ref{28}) and (\ref{29}) and define
$p_mf_0\in W^{(m)}$ to be the element of $W^{(m)}$ with the Fourier expansion 
(\ref{26}) and (\ref{27}).  Then the same estimates we
gave above imply that for $f_0\in W^{(0)}$ we have,
\[||\Delta p_mf_0||_{H^{\frac 12}({\cal E}_{\partial\Omega})}\le 
Cm||f_0||_{H^{\frac 12}({\cal E}_{\partial\Omega})},\]
for $\Delta p_m:=p_m-I$ and some constant $C$.  Now choose $m$ 
small enough so that the map,
\[P_0+P_0\Delta p_m:W^{(0)}\rightarrow W^{(0)},\]
is invertible.  Define,
\[g_0:=(P_0+P_0\Delta p_m)^{-1}f_0.\]
Then one can easily check that,
\[g_0+\Delta p_mg_0\in W^{(m)}\]
and,
\[P_0(g_0+\Delta p_mg_0)=f_0.\]
This shows that $P_0:W^{(m)}\rightarrow W^{(0)}$ is surjective.  
Furthermore if we define,
\[\delta =(I-P_0)\Delta p_m(P_0+P_0\Delta p_m)^{-1},\]
as a map from $W^{(0)}$ to $(I-P_0)H^{\frac 12}({\cal E}_{\partial
\Omega})$ then it is easy to
check that $f_0+\delta f_0\in W^{(m)}$ and $\delta$ satisfies the estimate
(2) of the theorem. QED
\section{Convergence Results}
In this section we provide the details for the 
approximation scheme for $\delta {\cal W}_j$ that was outlined
in \cite{P00}.  Let $f_j$ denote the section of ${\cal E}_{\Omega}$ defined in
(\ref{inhom}) above.  As a first approximation to $\delta {\cal W}_
j$ we
define,
\begin{equation}\delta_1{\cal W}_j=G_0(1+mG_0)^{-1}f_j\label{approx1}\end{equation}
We will show that for all sufficiently small $m$, $\delta_1{\cal W}_
j$ is
well defined and
\begin{equation}(m-\dsl)\delta_1{\cal W}_j=f_j,\label{31}\end{equation}
with $\delta_1{\cal W}_j|_{\partial\Omega}\in W^{(0)}$.  Thus $\delta_
1{\cal W}_j$ satisfies the same 
differential equation as $\delta {\cal W}_j$ but has boundary values
in $W^{(0)}$ instead of $W^{(m)}.$  Next we define,
\begin{equation}\delta_2{\cal W}_j=\delta (\delta_1{\cal W}_j).\label{approx2}\end{equation}
Then the boundary values of $\delta_1{\cal W}_j+\delta_2{\cal W}_
j$ are in $W^{(m)}$ but
since $\dsl\delta_2{\cal W}_j=0$ on $\Omega$ we find that,
\[(m-\dsl)(\delta_1{\cal W}_j+\delta_2{\cal W}_j)=f_j+m\delta_2{\cal W}_
j.\]
Now let $\delta_3{\cal W}_j$ denote the solution of, 
\begin{equation}(m-\dsl)\delta_3{\cal W}_j=-m\delta_2{\cal W}_j,\label{approx3}\end{equation}
such that $\delta_3{\cal W}_j|_{\partial\Omega}\in W^{(m)}.$   The solution we are 
interested in is then,
\[\delta {\cal W}_j=\delta_1{\cal W}_j+\delta_2{\cal W}_j+\delta_
3{\cal W}_j.\]
The following theorem gives convergence results that
will allow us to calculate (\ref{fcoef}) in the limit 
$m\rightarrow 0$.  We write $f_{0j}$ for the $m\rightarrow 0$ limit of $
f_j$,
\begin{equation}f_{0j}(z)=i\sqrt {\frac 2{\pi}}\left[\begin{array}{c}
-\bar {z}_j^{-\frac 12}\partial\varphi_j(z)\\
z_j^{-\frac 12}\bar{\partial}\varphi_j(z)\end{array}
\right].\label{31b}\end{equation}
\begin{theorem}\label{convergence}
For $\delta_k{\cal W}_j$ defined above $(k=1,2,3)$ we have,
\begin{list}{}{\setlength{\leftmargin}{.1in}}
\item{\bf 1.}  As $m\rightarrow 0$, $\delta_1{\cal W}_j$ converges to $
G_0f_{0j}$ in $W^{1,p}({\cal E}_{\Omega})$ for all 
$p>2$.
\item{\bf 2.} $||\delta_2{\cal W}_j||_{H^1({\cal E}_{\Omega})}\le 
Cm$ for some constant $C$ 
independent of $m$.
\item{\bf 3.} The Fourier coefficient,
\begin{equation}\int_{\theta_{{\bf r}}}^{\theta_{{\bf r}}+2\pi}(\delta_
3{\cal W}_j)_1(\epsilon e^{i\Theta_j})e^{-i\frac {\Theta_j}2}d\Theta_
j,\label{31a}\end{equation}
tends to 0 as $m\rightarrow 0$.
\end{list}

\end{theorem}

\emph{Remark}. As we shall see below, the upshot of 
these estimates is that we can compute the $m\rightarrow 0$ limit
of $mc^j_1({\cal W}_j)$ by calculating the appropriate Fourier 
coefficient of $G_0f_{0j}.$ Also in the course of proving 1-3 
of theorem (\ref{convergence}) we will confirm the
properties asserted for $\delta_k{\cal W}_j$, $k=1,2,3,$ when they were introduced
above.

\emph{Proof}. Estimate 1 of theorem (\ref{greenfunction}) 
shows that
for $p>2$, $G_0$ is bounded on $W^{1,p}({\cal E}_{\Omega})$.  Thus for small
enough $m$, the map $I+mG_0$ is invertible on $W^{1,p}({\cal E}_{
\Omega})$.  
Since it is clear that $f_j\in C^{\infty_{}}_0({\cal E}_{\Omega})
\subset W^{1,p}({\cal E}_{\Omega})$ it follows 
that $\delta_1{\cal W}_j\in W^{1,p}({\cal E}_{\Omega})$ for $p>2$.  Since $
\delta_1{\cal W}_j$ is in the 
image of $G_0$, part 3 of theorem (\ref{greenfunction}) implies
that the boundary value of $\delta_1{\cal W}_j$ on $\partial\Omega$ is in $
W^{(0)}.$ We
use part 2 of theorem (\ref{greenfunction}) to do the following 
calculation,
\begin{eqnarray*}
(m-\dsl)G_0(I+mG_0)^{-1}f_j&=&mG_0(I+mG_0)^{-1}f_j+(I+mG_0)^{-1}f_
j\\
&=&(mG_0+I)(I+mG_0)^{-1}f_j=f_j,\end{eqnarray*}
which confirms (\ref{31}).
Since $G_0$ is bounded on $W^{1,p}({\cal E}_{\Omega})$ the operator $
(I+mG_0)^{-1}$ 
converges uniformly to $I$ on $W^{1,p}({\cal E}_{\Omega})$ as $m\rightarrow 
0$.  Thus to finish
the proof of 1 we need only show that for all $p>2$ the
section $f_j$ converges in $W^{1,p}({\cal E}_{\Omega})$ to $f_{0j}
.$  Using 
(\ref{inhom}) one finds,
\begin{equation}f_j=i\sqrt m\left[\begin{array}{c}
e^{i\frac {\Theta_j}2}(I_{\frac 12}(mr)-I_{-\frac 12}(mr))\partial
\varphi_j(z)\\
e^{-i\frac {\Theta_j}2}(I_{-\frac 12}(mr)-I_{\frac 12}(mr))\bar{\partial}
\varphi_j(z)\end{array}
\right]\label{32}\end{equation}
The simple estimate,
\[I_{\pm\frac 12}(mr)=\left(\frac {mr}2\right)^{\pm\frac 12}\sum_{
n=0}^{\infty}\frac {(mr)^{2n}}{2^{2n}n!\Gamma (n\mp\frac 12)}\le\left
(\frac {mr}2\right)^{\pm\frac 12}\frac {\exp\left(\frac {2m\epsilon}
2\right)^2}{\Gamma\left(\frac 32\right)},\]
which is valid for $r<2\epsilon$ (which contains the support of
$\partial\varphi$$_j$ and $\bar{\partial}\varphi_j$) shows that dominated convergence applies
to,
\[\lim_{m\rightarrow 0}\int_{\Omega}|f_j-f_{0j}|^pidzd\bar {z}=0,\]
for all $p\ge 1$. The same estimate shows that 
dominated convergence applies to the $m\rightarrow 0$ limit of the 
integral,
\[\int_{\Omega}|df_j-df_{0j}|^pidzd\bar {z}\]
and this proves that $f_j$ converges to $f_{0j}$ in $W^{1,p}({\cal E}_{
\Omega})$.
Now fix $p>2$.  Since $f_j$ converges in $W^{1,p}({\cal E}_{\Omega}
)$ as
$m\rightarrow 0$ it follows that its norm in this space is uniformly
bounded.  Hence the $W^{1,p}$ norm of $\delta_1{\cal W}_j$ is also uniformly
bounded.  However since the domain $\Omega$ is bounded the 
$W^{1,p}({\cal E}_{\Omega})$ norm for $p>2$ dominates (a constant times) 
the $H^1({\cal E}_{\Omega})$ norm.  This shows that $\delta_1{\cal W}_
j$ is uniformly
bounded in $H^1({\cal E}_{\Omega})$ as $m\rightarrow 0$.  The Sobolev trace theorem 
\cite{LL97}
implies that the boundary value of $\delta_1{\cal W}_j$ is uniformly
bounded in $H^{\frac 12}({\cal E}_{\partial\Omega})$ and estimate 2 of theorem 
(\ref{isoproj}) then shows that $\delta_2{\cal W}_j=\delta (\delta_
1{\cal W}_j)$ has
norm in $H^{\frac 12}({\cal E}_{\Omega})$ bounded by $Cm$. Finally estimate (\ref{23})
shows that the extension of $\delta_2{\cal W}_j$ to $\Omega$ has $
H^1({\cal E}_{\Omega})$ norm
dominated by $Cm$, which is estimate 2 of theorem 
(\ref{convergence}).  

Before we turn to the proof of part
3 of theorem (\ref{convergence}) it will be useful to
establish the following estimates for solutions to
the massive Dirac equation.
\begin{theorem}\label{estimates}
Suppose that $\Psi\in L^2({\cal E}_{\Omega})$ is a weak solution to the 
Dirac equation,
\[(m-\dsl)\Psi =f,\]
in $\Omega ,$ where $f\in C^{\infty}_0({\cal E}_{\Omega})$.  Suppose that $
\Psi |_{\partial\Omega}\in W^{(m)}$
so that for $z\in C_{\epsilon}(a_j)$ the section $\Psi$ has the
local expansion,
\[\Psi (z)=\sum_{n\ge 0}c^j_n(\Psi )w_n(z_j)+c_n^j(\Psi )w_n^{*}(
z_j),\]
with $c_0^j(\Psi )=c_0^{j*}(\Psi )$, and for $z\in C_R$ the section
$\Psi$ has the local expansion,
\[\Psi (z)=\sum_{n\in {\bf Z}}c^{\infty}_n(\Psi )\hat {w}_n(z).\]
\[\]
Then,
\begin{equation}||\Psi ||_{L^2({\cal E}_{\Omega})}\le\frac 1m||f|
|_{L^2({\cal E}_{\Omega})},\label{33}\end{equation}
\begin{equation}\sum_j|c_0^j(\Psi )|^2\le\frac 18||f||_{L^2({\cal E}_{
\Omega})}^2\label{34}\end{equation}
\begin{equation}4\pi\sum_{n\ge 0}\left\{\left|c_n^j(\Psi )\right|^
2+\left|c_n^{j*}(\Psi )\right|^2\right\}I_{n+\frac 12}(m\epsilon 
)I_{n-\frac 12}(m\epsilon )\epsilon\le\frac 1m||f||^2_{L^2({\cal E}_{
\Omega})}\label{34a}\end{equation}
\begin{equation}4\pi\sum_{n\in {\bf Z}}|c^{\infty}_n(\Psi )|^2K_n
(mR)K_{n-1}(mR)R\le\frac 1m||f||^2_{L^2({\cal E}_{\Omega})}.\label{35}\end{equation}
\end{theorem}
\emph{Proof}. Since $f\in C_0^{\infty}({\cal E}_{\Omega})$, the existence theorem 
in \cite{P00} then gives us a weak solution
$\Psi\in L^2({\cal E}_{\Omega})$ of 
\[(m-\dsl)\Psi =f\]
which is smooth as a consequence of local
elliptic regularity.  Next we calculate the exterior
derivative of $2i\bar{\Psi}_1\Psi_2d\bar {z}$ using the fact that $
\Psi$ satisfies
the Dirac equation,
\[d(2i\bar{\Psi}_1\Psi_2d\bar {z})=m|\Psi |^2idzd\bar {z}-(\bar{\Psi}_
1f_1+\Psi_2\bar {f}_2)idzd\bar {z}.\]
Integrating this equality over $\Omega$ and using Stokes' 
theorem we find,
\begin{equation}m||\Psi ||^2_{L^2({\cal E}_{\Omega})}-2i\int_{\partial
\Omega}\bar{\Psi}_1\Psi_2d\bar {z}=\int_{\Omega}(\bar{\Psi}_1f_1+
\Psi_2\bar {f}_2)idzd\bar {z}.\label{36}\end{equation}
From this we deduce the inequality,
\begin{equation}m||\Psi ||^2_{L^2({\cal E}_{\Omega})}-2i\int_{\partial
\Omega}\bar{\Psi}_1\Psi_2d\bar {z}\le ||\Psi ||_{L^2({\cal E}_{\Omega}
)}||f||_{L^2({\cal E}_{\Omega})}\label{37}\end{equation}
Next we compute the boundary term in (\ref{37}) using
the local expansions for $\Psi$.  We find,
\begin{equation}-2i\int_{C_R}\bar{\Psi}_1\Psi_2d\bar {z}=4\pi\sum_
n|c_n^{\infty}(\Psi )|^2K_n(mR)K_{n-1}(mR)R\label{38}\end{equation}
and recalling the appropriate orientation of $\partial\Omega$,
\begin{eqnarray*}
2i\int_{C_{\epsilon}(a_j)}\bar{\Psi}_1\Psi_2d\bar {z}=4\pi\sum_{n
\ge 0}\left(|c_n^j(\Psi )|^2+|c_n^{j*}(\Psi )|^2\right)I_{n+\frac 
12}(m\epsilon )I_{n-\frac 12}(m\epsilon )\epsilon\\
+4\pi\bar {c}_0^j(\Psi )c_0^{j*}(\Psi )I_{-\frac 12}^2(m\epsilon 
)\epsilon +4\pi\bar {c}_0^{j*}(\Psi )c_0^j(\Psi )I_{\frac 12}^2(m
\epsilon )\epsilon .\end{eqnarray*}
The boundary condition $c_0^j(\Psi )=c_0^{j*}(\Psi )$ implies that the 
right hand side of this last equation is positive definite.
Thus the boundary term on the left hand side of (\ref{37}) 
is positive and we immediately deduce,
\[m||\Psi ||_{L^2({\cal E}_{\Omega})}^2\le ||\Psi ||_{L^2({\cal E}_{
\Omega})}||f||_{L^2({\cal E}_{\Omega})},\]
which is (\ref{33}).  Now (\ref{38}) and (\ref{37}) coupled
with the positivity of all the boundary contributions 
and (\ref{33}) together imply (\ref{34a}) and (\ref{35}).  
For the same
reasons we can pick out just one term from each of the
$C_{\epsilon}(a_j)$ boundary terms to find the inequality,
\[4\pi\sum_j|c_0^j(\Psi )|^2I_{-\frac 12}^2(m\epsilon )\epsilon\le\frac 
1m||f||_{L^2({\cal E}_{\Omega})}^2.\]
This must be true for all $\epsilon$ and since,
\[\lim_{\epsilon\rightarrow 0}I^2_{-\frac 12}(m\epsilon )\epsilon 
=\frac 2{m\pi},\]
we have proved (\ref{34}). QED 
(Theorem(\ref{estimates}))

Now suppose as in the preceeding theorem that $\Psi$ is an
$L^2({\cal E}_{\Omega})$ solution to,
\[(m-\dsl)\Psi =f,\]
where $f\in C^{\infty}_0({\cal E}_{\Omega}).$ To finish the proof of 3 of
Theorem(\ref{convergence}) we want to estimate $c_1^j(\Psi )$ in terms
of $f$.  To do this we first introduce the function,
\[V_j(z)=(z-a_j)^{-\frac 32}\prod_{k\ne j}(a_j-a_k)^{\frac 12}(z-
a_k)^{-\frac 12},\]
which we take to be branched along the rays ${\bf r}_j$.  Next
observe that if $\Psi$ is identified with its ${\cal U}_0$ trivialization
then $V_j\Psi$ is differentiable on $\Omega$ and,
\begin{eqnarray*}
d(2V_j\Psi_1dz)&=&-\bar{\partial }(2V_j\Psi_1)dzd\bar {z}=-2V_j\bar{
\partial}\Psi_1dzd\bar {z}\\
&=&mV_j\Psi_2dzd\bar {z}-V_jf_2dzd\bar {z}.\end{eqnarray*}
Stokes' theorem implies,
\[\int_{\partial\Omega}2V_j\Psi_1dz=\int_{\Omega}mV_j\Psi_2dzd\bar {
z}-\int_{\Omega}V_jf_2dzd\bar {z},\]
from which, together with (\ref{33}), we deduce the 
inequality,
\begin{equation}\left|\int_{\partial\Omega}V_j\Psi_1dz\right|\le 
||V_j||_{L^2(\Omega )}||f||_{L^2({\cal E}_{\Omega})}.\label{39}\end{equation}

Next we wish to estimate the boundary integrals over
$C_{\epsilon}(a_k)$.  First observe that in a neighborhood of $a_
k$ the
function $V_j$ has a ``Laurent'' expansion in powers of
$z_k=z-a_k,$
\[V_j(z)=\sum_{n\ge -1}c_n^k(V_j)z_k^{n-\frac 12},\]
and $z$ is restricted to an
annulus about $|z_k|=\epsilon$.   One finds,
\begin{eqnarray*}
\frac 1{2\pi\epsilon i}\int_{C_{\epsilon}(a_k)}V_j\Psi_1dz=c_1^k(
\Psi )c_{-1}^k(V_j)I_{\frac 12}(m\epsilon )\epsilon^{-\frac 32}+c_
0^k(\Psi )c_0^k(V_j)I_{-\frac 12}(m\epsilon )\epsilon^{-\frac 12}\\
+\sum_{n\ge 0}c_n^{k*}(\Psi )c_n^k(V_j)I_{n+\frac 12}(m\epsilon )
\epsilon^{n-\frac 12}.\end{eqnarray*}
Observe that the term with $c_1^k(\Psi )$ is present only for 
$k=j$ since one can easily check that,
\[c_{-1}^k(V_j)=\delta_{jk}.\]
Next we use the fact that the Taylor expansion of $V_j$ 
for $|z|\ge R$ has the form,
\[V_j(z)=\sum_{n<-1}c^{\infty}_n(V_j)z^n,\]
to calculate,
\[\frac 1{2\pi Ri}\int_{C_R}V_j\Psi_1(z)dz=-\sum_{n<-1}c_n^{\infty}
(V_j)R^nc_{n+1}^{\infty}(\Psi )K_{n+1}(mR).\]
Note: the coefficients $c_n^{\infty}(V_j)$ are zero for $n>-\frac 
N2-1$,
where $N$ is the number of branch points, but we will not 
need this.

\[\]
Next we use (\ref{34a}) and Cauchy's inequality for the 
$\ell^2$ norm with weight,
\[I_{n+\frac 12}(m\epsilon )I_{n-\frac 12}(m\epsilon )\epsilon ,\]
to estimate,
\begin{equation}\sum_{n\ge 0}|c_n^{k*}(\Psi )||c_n^k(V_j)|I_{n+\frac 
12}(m\epsilon )\epsilon\epsilon^{n-\frac 12}\le A_kB_k,\label{40}\end{equation}
where
\[A^2_k=\sum_{n\ge 0}|c_n^{k*}(\Psi )|^2I_{n+\frac 12}(m\epsilon 
)I_{n-\frac 12}(m\epsilon )\epsilon\]
and 
\[B^2_k=\sum_{n\ge 0}\left|\frac {c_n^k(V_j)\epsilon^{n-\frac 12}}{
I_{n-\frac 12}(m\epsilon )}\right|^2I_{n+\frac 12}(m\epsilon )I_{
n-\frac 12}(m\epsilon )\epsilon .\]
Combining this with (\ref{1}) we see that,
\[B^2_k\le\sum_{n\ge 0}\left|c_n^k(V_j)\epsilon^{n-\frac 12}\right
|^2\frac {m\epsilon^2}{n+\frac 12}\le\epsilon m||V_j||^2_{L^2(C_{
\epsilon}(a_k))},\]
since $c_n^k(V_j)\epsilon^{n-\frac 12}$ are Fourier coeffients for $
V_j$ on the 
circle $C_{\epsilon}(a_k)$.  The norm of $V_j$ that appears here is actually 
the $H^{-\frac 12}$ norm, 
but this won't matter for us.  This last estimate for $B_k$ 
combined with (\ref{34a}) for $A_k$ give us,
\begin{equation}A_kB_k\le\sqrt {\frac {\epsilon}{4\pi}}||V_j||_{L^
2(C_{\epsilon}(a_k))}||f||_{L^2({\cal E}_{\Omega})}.\label{41}\end{equation}
In a similar fashion we estimate,
\begin{equation}\sum_{n<-1}|c_n^{\infty}(V_j)R^n||c_{n+1}^{\infty}
(\Psi )|K_{n+1}(mR)R\le A_{\infty}B_{\infty},\label{42}\end{equation}
where,
\[A_{\infty}^2=\sum_{n<-1}|c_{n+1}^{\infty}(\Psi )|^2K_{n+1}(mR)K_
n(mR)R\le\frac 1{4\pi m}||f||_{L^2({\cal E}_{\Omega})}^2,\]
and
\[B_{\infty}^2=\sum_{n<-1}|c_n^{\infty}(V_j)R^n|^2\frac {K_{n+1}(
mR)}{K_n(mR)}R.\]
In this equation it is important for us that $n<-1$.  For
$n<-1$ we have,
\[\frac {K_{n+1}(mR)}{K_n(mR)}=\frac {K_{|n|-1}(mR)}{K_{|n|}(mR)}
\le\frac {mR}{|n|-1},\]
so that
\[B^2_{\infty}\le mR||V_j||^2_{L^2(C_R)},\]
and 
\begin{equation}A_{\infty}B_{\infty}\le\sqrt {\frac R{4\pi}}||V_j
||_{L^2(C_R)}||f||_{L^2({\cal E}_{\Omega})}.\label{43}\end{equation}
Combining the expressions we found for the boundary 
integrals with the estimates that follow from 
(\ref{40})--(\ref{43}) we find the following lower bound,
\begin{eqnarray*}
\frac 1{2\pi}\left|\int_{\partial\Omega}V_j\Psi_1dz\right|\ge |c_
1^j(\Psi )|I_{\frac 12}(m\epsilon )\epsilon^{-\frac 12}-\sum_k|c_
0^k(V_j)||c_0^k(\Psi )|I_{-\frac 12}(m\epsilon )\epsilon^{\frac 1
2}\\
-C||V_j||_{L^2(\partial\Omega )}||f||_{L^2({\cal E}_{\Omega})},\end{eqnarray*}
for a constant $C$ which is independent of $m$.
Now we put together this lower bound with (\ref{39}) to
find,
\begin{equation}I_{\frac 12}(m\epsilon )|c_1^j(\Psi )|\le C(I_{-\frac 
12}(m\epsilon )\sum_k|c_0^k(\Psi )|+||f||_{L^2({\cal E}_{\Omega})}
),\label{44}\end{equation}
where the constant $C$ is independent of $m$ but 
incorporates all the dependence on $V_j$. The
form of this estimate that we will use is now obtained
by combining (\ref{34}) with (\ref{44}).  We find,
\begin{equation}I_{\frac 12}(m\epsilon )|c_1^j(\Psi )|\le C(I_{-\frac 
12}(m\epsilon )+1)||f||_{L^2({\cal E}_{\Omega})},\label{45}\end{equation}
for a different constant $C$.

We are now prepared to finish the proof of part 3 of
Theorem (\ref{convergence}).  Recall that $\delta_3{\cal W}_j$ is defined
as the solution to,
\begin{equation}(m-\dsl)\delta_3{\cal W}_j=-m\delta_2{\cal W}_j,\label{48}\end{equation}
with boundary values in $W^{(m)}$.  We could make this 
description technically precise and prove the existence
of such a solution using $H^1$ estimates along the lines
of the $L^2$ estimate (\ref{33}).  The relevant estimates
can be obtained via a Stokes' theorem calculation 
involving the exterior derivative,
\[2id\left((\bar\Psi_1\bar\partial\Psi_1+\bar\Psi_2\bar\partial\Psi_
2)d\bar z-(\bar\Psi_1\partial\Psi_1+\bar\Psi_2\partial\Psi_2)dz\right
),\]
for a solution $\Psi$ to
\[(m-\dsl)\Psi =f.\]
However it will be simpler to proceed differently.  The
solution of (\ref{48}) which is relevant to us is the one
constructed via functional analysis in \cite{P00}.  This solution
is a weak $L^2$ solution to (\ref{48}) inside $\Omega$ which extends 
to a solution of the homogeneous equation 
$(m-\dsl)\delta_3{\cal W}_j=0$ outside $\Omega$ and is globally in $
L^2({\cal E})$.  We
can obtain such a solution by approximating the right
hand side $\delta_2{\cal W}_j$ in $L^2({\cal E}_{\Omega})$ by a sequence of functions $
f_n\in C_0^{\infty}({\cal E}_{\Omega})$.
Theorem (\ref{estimates}) shows that the resulting 
sequence of solutions to the approximate equations 
tends strongly in $L^2({\cal E}_{\Omega})$  
to a weak solution to (\ref{48}). 
The norms on the left hand sides of (\ref{34a}) and 
(\ref{35}) are equivalent to the $L^2$ norms of limiting 
solution 
in the
components of the exterior of $\Omega$ and so the resulting
solution is globally in $L^2({\cal E})$ (this is a consequence of the
same Stokes' theorem calculation that went into the 
proof of Theorem (\ref{estimates}) but done in the 
components of the exterior of $\Omega$).  We may thus
identify the limiting solution with $\delta_3{\cal W}_j$ and by obtaining the
solution $\delta_3{\cal W}_j$ in this fashion we see that estimate
(\ref{45}) remains valid, so that,
\[I_{\frac 12}(m\epsilon )\left|c_1^j(\delta_3{\cal W}_j)\right|\le 
C(I_{-\frac 12}(m\epsilon )+1)m||\delta_2{\cal W}_j||_{L^2(\Omega 
)}.\]
The left hand side of this last inequality is the Fourier
coefficient (\ref{31a}) and the right hand side tends to
0 using estimate 2 of Theorem (\ref{convergence}).  This
finishes the proof of Theorem (\ref{convergence}). QED

\emph{Remark}. The norms on the left hand side of 
(\ref{34a}) and (\ref{35}) which are the 
appropriate norms for boundary values of $L^2$ solutions
also appear to be equivalent to the $H^{-\frac 12}$ norms on the 
corresponding circles.  The loss of one half derivative 
for boundary values of solutions to $(m=0)$ Dirac equations is
a general property \cite{BW93}.  In our case, this would follow from
the following Bessel function estimate,
\[\frac {K_n(r)}{K_{n-1}(r)}\le 2(n-1)\left(1+\frac 1r\right),\]
for $r>0$ and $n\ge 2$ which seems to be true.

We now substitute $\delta {\cal W}_j=\delta_1{\cal W}_j+\delta_2{\cal W}_
j+\delta_3{\cal W}_j$ into 
(\ref{fcoef}) and use Theorem (\ref{convergence}) to
determine the limit as $m\rightarrow 0$.  According to part 1
of Theorem (\ref{convergence}), $\delta_1{\cal W}_j$ converges to
$G_0f_{0j}$ in $W^{1,p}({\cal E}_{\Omega})$ for $p>2$.  The Sobolev trace theorem
implies that it converges in $W^{\frac 12,p}({\cal E}_{\partial\Omega}
)$ and this is enough
to show that the Fourier coefficient of $\delta_1{\cal W}_j$ which 
appears in (\ref{fcoef}) converges to, 
\begin{equation}\sqrt {\frac {\pi}{2\epsilon}}\frac 1{2\pi}\int_{
\theta_{{\bf r}}}^{\theta_{{\bf r}}+2\pi}(G_0f_{0,j})_1(\epsilon 
e^{i\Theta_j})e^{-i\frac {\Theta_j}2}d\Theta_j.\label{49}\end{equation}
Estimate 2 of Theorem (\ref{convergence}) implies that
the $H^1({\cal E}_{\Omega})$ norm of $\delta_2{\cal W}_j$ tends to 0 as $
m\rightarrow 0$ and again
the Sobolev trace theorem implies that the boundary 
value tends to 0 in $H^{\frac 12}({\cal E}_{\partial\Omega})$ which is enough to show
that the contribution made by $\delta_2{\cal W}_j$ to (\ref{fcoef}) is 0 
in this limit.  Finally part 3 of Theorem (\ref{convergence})
shows that $\delta_3{\cal W}_j$ makes no contribution to the 
$m\rightarrow 0$ limit of (\ref{fcoef}).  Thus to finish the proof of
Theorem (\ref{main}) we need only calculate (\ref{49}).

We turn now to the calculation of (\ref{49}).  Using the
definition of $f_{0,j}$ found in (\ref{31b}) above we see that,
\[(G_0f_{0,j})_1(z)=i\sqrt {\frac 2{\pi}}\int_{\Omega}\left(-G_{1
1}(z,z')\bar z_j^{\prime -\frac 12}\partial\varphi (z')+G_{12}(z,
z')z^{\prime -\frac 12}_j\bar\partial\varphi\right)dz'd\bar {z}'.\]
Using the fact that $G_{11}(z,z')$ is anti-holomorphic in $z'$ 
and $G_{12}(z,z')$ is holomorphic in $z$ we can rewrite this 
last integral as the integral of an exact form,
\[i\sqrt {\frac 2{\pi}}\int_{\Omega}d\left(-G_{11}(z,z')\bar z_j^{
\prime -\frac 12}\varphi (z')d\bar z'-G_{12}(z,z')z^{\prime -\frac 
12}_j\varphi (z')dz'\right).\]
Since $\varphi (z)=1$ on $C_{\epsilon}(a_j)$ and vanishes on the rest of
$\partial\Omega$, Stokes' theorem implies that the last integral is,
\[i\sqrt {\frac 2{\pi}}\int_{C_{\epsilon}(a_j)}G_{11}(z,z')\bar {
z}_j^{\prime -\frac 12}d\bar {z}'+G_{12}(z,z')z_j^{\prime -\frac 
12}dz'.\]
Now substitute, 
\[G_{11}(z,z')=-\frac 1{4\pi i}\sum_kv_k(z)\bar {u}_k(z),\]
and (\ref{13}) for $G_{12}(z,z')$ in this last integral and 
use the residue calculations that are found in 
the results that follow
(\ref{proj1}) to find,
\[(G_0f_{0,j})_1(z)=-i\frac {z_j^{-\frac 12}}{\sqrt {2\pi}}+i\sqrt {\frac 
2{\pi}}v_j(z).\]
Using this result it is now a simple matter to convert
the Fourier integral (\ref{49}) into the following residue
calculation,
\[\frac 1{2\pi}\int_{C_{\epsilon}(a_j)}z_j^{-2}\sum_{{|
\epsilon |=0},{\epsilon_j=\frac 12}}c(\epsilon )\prod_{k\ne j}\frac {
(z-a_k)^{\epsilon_k}}{(a_j-a_k)^{\epsilon_k}}\,\,dz\]
which in turn gives,
\[i\frac {\partial}{\partial z}\sum_{{|\epsilon |=0},{\epsilon_
j=\frac 12}}c(\epsilon )\prod_{k\ne j}\frac {(z-a_k)^{\epsilon_k}}{
(a_j-a_k)^{\epsilon_k}}\left|_{z=a_j}\right.=i\sum_{{|\epsilon 
|=0},{\epsilon_j=\frac 12}}c(\epsilon )\sum_{k\ne j}\frac {\epsilon_
k}{a_j-a_k}.\]
Dividing by $2i$ to get the limiting value of the 
coefficient that appears in the $m\rightarrow 0$ of the log
derivative of the tau function we find,
\begin{equation}\lim_{m\rightarrow 0}\frac {mc_1^j({\cal W}_j)}{2
i}=\frac 12\sum_{{|\epsilon |=0},{\epsilon_j=\frac 12}}c
(\epsilon )\sum_{k\ne j}\frac {\epsilon_k}{a_j-a_k}.\label{50}\end{equation}
To finish the proof of Theorem (\ref{main}) we compare 
this result with,
\begin{equation}\frac {\partial}{\partial a_j}\sum_{|\epsilon |=0}
\prod_{\alpha <\beta}|a_{\alpha}-a_{\beta}|^{2\epsilon_{\alpha}\epsilon_{
\beta}}=\sum_{|\epsilon |=0}\sum_{k\ne j}\frac {\epsilon_j\epsilon_
k}{a_j-a_k}\prod_{\alpha <\beta}|a_{\alpha}-a_{\beta}|^{2\epsilon_{
\alpha}\epsilon_{\beta}}\label{51}\end{equation}
which we obtained using,
\[\frac {\partial}{\partial a_j}|a_{\alpha}-a_{\beta}|^{2\epsilon_{
\alpha}\epsilon_{\beta}}=\frac {\epsilon_{\alpha}\epsilon_{\beta}}{
a_{\alpha}-a_{\beta}}(\delta_{\alpha j}-\delta_{\beta j})|a_{\alpha}
-a_{\beta}|^{2\epsilon_{\alpha}\epsilon_{\beta}}.\]
In (\ref{50}) observe that the $|\epsilon |=0$ sum has two 
different possible values for $\epsilon_j$, either $\epsilon_j=\frac 
12$ or 
$\epsilon_j=-\frac 12$.  However since the summand on the right hand
side of (\ref{50}) is clearly invariant under the complete
sign reversal $\epsilon_{\alpha}$$\rightarrow -\epsilon_{\alpha}$ it follows that the whole sum is
just twice the result for $\epsilon_j=\frac 12$.  That is,
\[\frac {\partial}{\partial a_j}\sum_{|\epsilon |=0}\prod_{\alpha 
<\beta}|a_{\alpha}-a_{\beta}|^{2\epsilon_{\alpha}\epsilon_{\beta}}
=\sum_{{|\epsilon |=0},{\epsilon_j=\frac 12}}\sum_{k\ne 
j}\frac {\epsilon_k}{a_j-a_k}\prod_{\alpha <\beta}|a_{\alpha}-a_{
\beta}|^{2\epsilon_{\alpha}\epsilon_{\beta}}.\]
Comparing this with (\ref{50}) and recalling the 
definition of $c(\epsilon )$ we have finished the proof of Theorem
(\ref{main}).
\section{Odd Correlations and Holonomic Fields}
In this section we make some observations about the 
application of the technique used to prove 
Theorem(\ref{main}) to work out the asymptotics of the
odd Ising scaling functions from below $T_c$ and also the
short distance behavior of the correlations for Holonomic
Quantum Fields.  

First we treat the case where $n$ is odd.  
The one difference in the analogue of Lemma(\ref{pchar}) 
for $n$ odd is
that the subspace $W^{(m)}_{\infty}$ is now the $L^2$ closure of the span
of, 
\[\hat {w}_n(z)=\left[\begin{array}{c}
-e^{-i(n+\frac 12)\theta}K_{n+\frac 12}(m|z|)\\
e^{-i(n-\frac 12)\theta}K_{n-\frac 12}(m|z|)\end{array}
\right],\]
for $n\in {\bf Z}$.  For definiteness we make the choice 
$0<\theta <2\pi$ and choose the ${\cal U}_{\infty}$ trivialization (in
the complement of $\theta =0$) so that finite linear 
combinations of the $\hat {w}_n(z)$  are smooth sections
of ${\cal E}$ in the ${\cal U}_{\infty}$ trivialization.  Without difficulty
one can compute the $m\rightarrow 0$ limit of the normalized
versions of these vectors and as a consequence
we define $W^{(0)}_{\infty}$ as the $L^2$ closure of the span
of 
\[\left\{\left[\begin{array}{c}
e^{-i(n+\frac 12)\theta}\\
0\end{array}
\right]\right\}_{n\ge 1},\left[\begin{array}{c}
-e^{-i\frac {\theta}2}\\
e^{i\frac {\theta}2}\end{array}
\right],\left\{\left[\begin{array}{c}
0\\
e^{i(n+\frac 12)\theta}\end{array}
\right]\right\}_{n\ge 1}.\]

Next we introduce a
Green function $-\dsl$ with $W^{(0)}$ boundary conditions 
in the following manner.
\begin{equation}G_0(z,z')=-\frac 1{4\pi i}\left[\begin{array}{ccc}
\sum_ju_j(z)\overline {v_j(z')}&g(z,z')\\
\overline {g(z,z')}&\sum_j\overline {u_j(z)}v_j(z')&\end{array}
\right],\label{51}\end{equation}
where,
\begin{equation}u_j(z):=(z-a_j)^{-\frac 12}\prod_{k\ne j}\frac {(
z-a_k)^{\frac 12}}{(a_j-a_k)^{\frac 12}},\label{52}\end{equation}
\begin{equation}g(z,z'):=\sum_{|\epsilon |=\pm\frac 12}c(\epsilon 
)\frac {\prod_j(z-a_j)^{\epsilon_j}(z'-a_j)^{-\epsilon_j}}{z'-z},\label{53}\end{equation}
with $\epsilon =(\epsilon_1,\ldots ,\epsilon_N)$ and each $\epsilon_
j=\pm\frac 12$.  Also
\[|\epsilon |:=\sum_{j=1}^N\epsilon_j,\]
\begin{equation}c(\epsilon ):=\frac {\prod_{j<k}|a_j-a_k|^{2\epsilon_
j\epsilon_k}}{\sum_{|\epsilon |=\pm\frac 12}\prod_{j<k}|a_j-a_k|^{
2\epsilon_j\epsilon_k}},\label{54}\end{equation}
and
\begin{equation}v_j(z)=(z-a_j)^{-\frac 12}\sum_{|\epsilon |=\pm\frac 
12,\epsilon_j=\frac 12}c(\epsilon )\prod_{k\ne j}\frac {(z-a_k)^{
\epsilon_k}}{(a_j-a_k)^{\epsilon_k}}.\label{55}\end{equation}
  The multivalued functions
$(z-a_j)^{\epsilon_j}$ are all defined using
the argument $\Theta_j$ and are consequently branched along 
$z\in {\bf r}_j$ .  We regard $G_0(z,z')$ as defining an operator, $
G_0,$
acting on sections of ${\cal E}_{\Omega}$ in the following manner,
\begin{equation}G_0f(z):=\int_{\Omega}G_0(z,z')f(z')dz'd\bar {z}'
,\label{56}\end{equation}
where the section $f(z')$ is identified with its ${\cal U}_0$ 
trivialization.  We also regard $G_0f$ as a section of
${\cal E}_{\Omega}$ given in the ${\cal U}_0$ trivialization.  

The homogeneous function identity, 
\[\sum_k\bar {u}_k(z)v_k(z')=\sum_k\bar {v}_k(z)u_k(z'),\]
can be proved along the lines of Lemma(\ref{homo}) and
this makes it possible to establish the desired results
concerning the Green function and the projection $P_0$.
One matter that requires a little further analysis is 
the proof that that $G_0f$ has boundary values on $C_R$ 
which are in $W^{(0)}_{\infty}$.  For this purpose it is useful to
introduce a ${\cal U}_{\infty}$ trivialization for ${\cal E}$ 
over
$\{z:|z|>R\}\backslash \{t\in {\bf R}:t>0\}$ by introducing
square root $z^{\frac 12}=|z|^{^{\frac 12}}e^{i\frac {\theta}2}$  for $
0<\theta <2\pi$, which is branched 
along the positive real axis. Smooth sections of ${\cal E}$ over
$\{z:|z|>R\}\backslash \{t\in {\bf R}:t>0\}$ can then be represented in the 
${\cal U}_{\infty}$ trivialization as products $z^{\frac 12}\phi 
(z)$ for a smooth map $\phi$ from
$D_{\infty}:=\{z:|z|>R\}$ into ${\bf C}^2$.  For the purpose of analysing the 
behavior of the Green function $G_0(z,z')$ for $|z|>R$ it is
useful to note that $v_j(z)$ has a representation in this
domain given by,
\begin{eqnarray*}
v_j(z)=z^{-\frac 12}\left(1-\frac {a_j}z\right)^{-\frac 12}\sum_{
|\epsilon |=\frac 12,\epsilon_j=\frac 12}c(\epsilon )\prod_{k\ne 
j}\frac {\left(1-\frac {a_k}z\right)^{\epsilon_k}}{(a_j-a_k)^{\epsilon_
k}}\\
+z^{-\frac 32}\left(1-\frac {a_j}z\right)^{-\frac 12}\sum_{|\epsilon 
|=-\frac 12,\epsilon_j=\frac 12}c(\epsilon )\prod_{k\ne j}\frac {\left
(1-\frac {a_k}z\right)^{\epsilon_k}}{(a_j-a_k)^{\epsilon_k}}.\end{eqnarray*}
Using this one can check that for $\varphi\in C^{\infty}_0({\cal E}_{
\Omega})$ we have
$G_0\varphi |_{C_R}\in W^{(0)}_{\infty}$ provided the following reality conditions
are satisfied,
\[\sum_{|\epsilon |=\frac 12,\epsilon_j=\frac 12}c(\epsilon )\prod_{
k\ne j}(a_j-a_k)^{-\epsilon_k}=\sum_{|\epsilon |=\frac 12,\epsilon_
j=\frac 12}\bar {c}(\epsilon )\prod_{k\ne j}\overline {(a_j-a_k)}^{
-\epsilon_k}.\]
This will be true for our choice of $c(\epsilon )$ provided that,
\[\sum_{|\epsilon |=\frac 12,\epsilon_j=\frac 12}\prod_{{{\alpha 
<\beta}\atop {\alpha ,\beta\ne j}}}|a_{\alpha}-a_{\beta}|^{2\epsilon_{
\alpha}\epsilon_{\beta}}\prod_{k\ne j}|a_j-a_k|^{\epsilon_k}(a_j-
a_k)^{-\epsilon_k},\]
is real.  However, under the transformation 
$\epsilon_k\rightarrow -\epsilon_k$ for $k\ne j$ the product,
\[\prod_{k\ne j}|a_j-a_k|^{\epsilon_k}(a_j-a_k)^{-\epsilon_k},\]
maps into its complex conjugate while in the preceeding
sum the coefficient of this product is real and invariant.
This implies reality for the sum.  

The rest of the analysis closely follows that in the even case 
and so we will only quote the final result.  For $N$ odd
we have,
\[\lim_{m\rightarrow 0}d_a\log\tau_{-}(ma)=\frac 12d_a\log\sum_{|
\epsilon |=\pm\frac 12}\prod_{\alpha <\beta}|a_{\alpha}-a_{\beta}
|^{2\epsilon_{\alpha}\epsilon_{\beta}}.\]

Finally we describe the situation for the tau functions
for holonomic fields in the formalism of \cite{P93}.  
Suppose that for $j=1,\ldots ,N$ we have,
\[-\frac 12<\lambda_j<\frac 12,\]
and for simplicity we also suppose that,
\[\sum_j\lambda_j=0.\]
The restricted local expansion that determines the subspace $W^{(
m)}_j$ 
is,
\[w(z)=\sum_{{{k\in {\bf Z}+\frac 12}\atop {k>0}}}a_k^j(w)w_{k-\lambda_
j}(z_j)+b^j_k(w)w^{*}_{k+\lambda_j}(z_j).\]
At infinity the restricted expansion that determines 
$W^{(m)}_{\infty}$ is,
\[w(z)=\sum_{k\in {\bf Z}+\frac 12}c_k(w)\hat {w}_k(z).\]
Without difficulty one can check that the limiting 
subspaces, $W^{(0)}_j$, are spanned by,
\[\left[\begin{array}{c}
z_j^{k-\lambda_j}\\
0\end{array}
\right],\left[\begin{array}{c}
0\\
\bar {z}_j^{k+\lambda_j}\end{array}
\right]\mbox{\rm \ for }k=\frac 12,\frac 32,\ldots\]
and $W^{(0)}_{\infty}$ is spanned by,
\[\left[\begin{array}{c}
z^{-n}\\
0\end{array}
\right],\left[\begin{array}{c}
0\\
\bar {z}^{-n}\end{array}
\right]\mbox{\rm \ for }n=1,2,3\ldots\]
The mass 0 Green function for the Dirac operator
of interest is clearly (see Proposition 1.1 in \cite{P93}),
\[G_0(z,z')=-\frac 1{4\pi}\left[\begin{array}{cc}
0&g(z,z')\\
\overline {g(z,z')}&0\end{array}
\right],\]
where
\[g(z,z')=\frac {\prod_jz_j^{-\lambda_j}(z'_j)^{\lambda_j}}{z'-z}
.\]
There are no ``chiral symmetry breaking'' terms.  
In the notation of (4.3) of \cite{P93} we have,
\[d_a\log\tau (ma,\lambda )=\frac m2\sum_j\left\{a_{\frac 12,j}^j
(-\lambda )da_j+\bar a_{\frac 12,j}^j(\lambda )d\bar a_j\right\}.\]
and we find for the $m\rightarrow 0$ limit,
\[\lim_{m\rightarrow 0}d_a\log\tau (ma,\lambda )=\sum_{j=1}^N\left
\{\sum_{k\ne j}\frac {\lambda_j\lambda_k}{a_j-a_k}da_j+\sum_{k\ne 
j}\frac {\lambda_j\lambda_k}{\bar {a}_j-\bar {a}_k}d\bar a_j\right
\},\]
which is also just,
\[d_a\log\prod_{j<k}|a_j-a_k|^{\lambda_j\lambda_k}.\]
\emph{Acknowlegements}. The author would like to thank
Doug Pickrell for many helpful conversations.  He would
also like to thank Alexander Its for helpful remarks and
encouragement.

\end{document}